\documentclass[12pt]{article}
\usepackage{amssymb,amsmath,amsthm,amsfonts,amscd}
 \usepackage{graphicx}
\newcommand\be{\begin{equation}}
\newcommand\ee{\end{equation}}
\newcommand\bea{\begin{eqnarray}}
\newcommand\eea{\end{eqnarray}}
\newcommand\ket[1]{|#1\rangle}

\newcommand{\fatalpha}{{\bf \alpha \kern -0.44em \alpha}}
\newcommand{\fatsigma}{{\bf \sigma \kern -0.54em \sigma}}
\newcommand{\tpchi}{{\bf \chi \kern -0.35em \chi}}
\newcommand{\llambda}{{\bf \lambda \kern -0.45em \lambda}}

\renewcommand{\theequation}{\arabic{equation}}
\renewcommand{\theequation}{\thesection.\arabic{equation}}
\bibliography{plain}
\title{\bf Bell-states diagonal entanglement witnesses for relativistic and non-relativistic  multispinor systems in arbitrary
dimensions}\vspace{20mm}
\author{ M. A. Jafarizadeh$^{a,b,c}$
 \thanks{E-mail:jafarizadeh@tabrizu.ac.ir}  ,
 R. Sufiani  $^{a,b}$
 \thanks{E-mail:sofiani@tabrizu.ac.ir}
\\ $^a${\small Department of Theoretical Physics and Astrophysics,
University of Tabriz, Tabriz 51664, Iran.} \\ $^b${\small
Institute for Studies in Theoretical Physics and Mathematics,
Tehran 19395-1795, Iran.} \\ $^c${\small Research Institute for
Fundamental Sciences, Tabriz 51664, Iran.}} \pagebreak

\vspace{20mm}
\begin{document}
\maketitle \vspace{15mm}
\newpage
\begin{abstract}
Two kinds of Bell-states diagonal (BSD) entanglement witnesses (EW)
are constructed by using the algebra of Dirac $\gamma$ matrices in
the space-time of arbitrary dimension $d$, where the first kind can
detect some BSD relativistic and non-relativistic $m$-partite
multispinor bound entangled states in Hilbert space of dimension
$2^{m\lfloor d/2\rfloor}$, including the bipartite Bell-type and
iso-concurrence type states in the four-dimensional space-time
($d=4$). By using the connection between Hilbert-Schmidt measure and
the optimal EWs associated with states, it is shown that as far as
the spin quantum correlations is concerned, the amount of
entanglement is not a relativistic scalar and has no invariant
meaning. The introduced EWs are manipulated via the linear
programming (LP) which can be solved exactly by using simplex
method. The decomposability or non-decomposability of these EWs is
investigated, where the region of non-decomposable EWs of the first
kind is partially determined and it is shown that, all of the EWs of
the second kind are decomposable. These EWs have the preference that
in the bipartite systems, they can determine the region of separable
states, i.e., bipartite non-detectable density matrices of the same
type as the EWs of the first kind are necessarily separable. Also,
multispinor EWs with non-polygon feasible regions are provided,
where the problem is solved by approximate LP, and in contrary to
the exactly manipulatable EWs, both the first and second kind of the
optimal approximate EWs can detect some bound entangled states.


{\bf PACs Index: 03.65.Ud}
\end{abstract}
\vspace{70mm}
\newpage
\section{Introduction}
Entanglement is one of the most fascinating features of quantum
mechanics and a lot of work has been devoted to this topic in the
recent years \cite{lew}-\cite{brus}. It has recently been
recognized that entanglement is a very important resource in
quantum information processing \cite{2} such as teleportation
\cite{telep} and clock synchronization \cite{cync}. On the other
hand, there is a natural interest in studying nonlocal quantum
correlations in the framework of special relativity \cite{ryder}.
Relativistic quantum information processing is of growing interest
not only for the logical completeness but also with regard to new
features, such as the physical bounds on information transfer,
processing and the errors provided by the full relativistic
treatments (see the review \cite{1'}). Tracing back to Bell's
famous re-imagining of the Einstein-Podolosky-Rosen paradox
\cite{1}, a standard system of interest is two particles with
spins entangled due to their production in the decay or
scattering. Various authors have considered the entanglement of
two relativistic particles \cite{2'}-\cite{19'}. Some of these
papers discuss the covariance of the Bell's inequality and show
that the violation of this inequality decreases with increasing
the velocity of the moving frame. Although the results of this
type produce interesting insights to the relativistic quantum
information, but it should be noticed that decreasing the amount
of violation of the Bell inequality do not imply that the amount
of entanglement decreases under the Lorentz transformation, since
the violation of Bell inequalities are tools only for detection of
non-locality and can not be considered as a suitable entanglement
measure. On the other hand, these papers have studied only pure
relativistic states where, the entanglement between spins of two
electrons is considered. In this paper, we take the approach of
so-called entanglement witnesses (EW's) \cite{4'} to distinguish
separable mixed states from entangled ones (an EW for a given
entangled state $\rho$ is an observable $W$ whose expectation
value is non-negative on any separable state, but strictly
negative on the entangled state $\rho$) and by constructing EWs
called Bell-states diagonal (BSD) multispinor EWs, present a
general scheme which can be used for studying the entanglement
properties of relativistic and non-relativistic multispinor
systems in an arbitrary space-time dimension $d$. It should be
noticed that, the framework of Bell inequalities fits in the
scheme of EWs such that as it has been discussed in Ref.
\cite{witbel}, each Bell inequality can be viewed as a particular
example of an entanglement witness. In fact, the Bell inequalities
are corresponded to non-optimal EWs and can be only used as
criteria for detection of entanglement. Despite of the fact that
the EWs are designed mainly for detection of the entanglement, it
has been shown \cite{witmeas} that the optimal EW associated with
a density matrix $\rho$ -in the sense that, the expectation value
of the optimal EW (associated with $\rho$) over $\rho$ is the most
negative value between the expectation values of other EWs over
$\rho$- can be used as measure of entanglement quantifying the
amount of entanglement of $\rho$. In Refs. \cite{hilbersh},
\cite{16} a connection between Hilbert-Schmidt measure and the
optimal EW associated with a state has been discussed. We will use
this connection in order to show that, the amount of entanglement
between the spins of electron and positron (for a given momentum
$\vec{p}$) in a bipartite system with space-time dimension $d=4$
is not Lorentz invariant, where this result is in agreement with
those of Ref. \cite{peres'}. There has been much work on the
separability problem, particularly from the Innsbruck-Hannover
group, as reviewed in \cite{11,12}, which emphasizes convexity and
proceeds by characterizing EWs in terms of their extreme points,
the so-called optimal EWs \cite{13}, and PPT entangled states
(those density matrices which have positive partial transposition
with respect to each subsystem \cite{Lewen3}) in terms of their
extreme points, the edge PPT entangled states \cite{14,15}. In
fact, in order to a hermitian operator $W$ be an EW, it must
posses at least one negative eigenvalue and the expectation value
of $W$ over any separable state must be non-negative. Therefore,
for determination of EWs, one needs to determine the minimum value
of this expectation value over the feasible region (the minimum
value must be non-negative) and hence the problem reduces to an
optimization over the convex set of feasible region. For example,
in \cite{reza,reza1} the manipulation of generic Bell-state
diagonal EWs has been reduced to such an optimization problem. It
has been shown that, if the feasible region for this optimization
constructs a polygon by itself, the corresponding boundary points
of the convex hull will minimize exactly the optimization problem.
This problem is called linear programming (LP) and the simplex
method is the easiest way of solving it \cite{Boyd}. If the
feasible region is not a polygon, with the help of tangent planes
in this region at points which are determined either analytically
or numerically, one can define a new convex hull which is a
polygon and has encircled the feasible region. The points on the
boundary of the polygon can approximately determine the minimum
value of the optimization problem. Thus the approximated value is
obtained via LP. In general, it is difficult to find this region
and solve the corresponding optimization problem; thus, it is
difficult to find any generic multipartite EW. Recently, in Ref.
\cite{jsn}, a new class of EWs called reduction type EWs has been
introduced for which the feasible regions turn out to be convex
polygons. In this work, we construct two kinds of BSD multispinor
EWs by using the algebra of Dirac $\gamma$ matrices in the
space-time of arbitrary dimension $d$, where the first kind can
detect some $m$-partite BSD non-relativistic multispinor PPT
entangeled states with Hilbert space of dimension $2^{m\lfloor
d/2\rfloor}$. Furthermore, in the four-dimensional space-time
($d=4$), we introduce $16$ Bell-type and iso-concurrence type
states (for definition and entanglement properties of the
iso-concurrence states, the reader is refered to
\cite{akhtar}-\cite{rez}) and show that, these states (including
the spinor ``EPR" state \cite{pachos} which is a special kind of
iso-concurrence type entangled states) are detected by the
constructed multispinor EWs. Moreover, by using the bipartite
optimal EWs of the first kind and the Hilbert-Schmidt measure of
entanglement, we calculate the amount of entanglement for some
kinds of BSD density matrices (in the four-dimensional space-time)
in the rest frame and the corresponding Lorentz transformed
states, where the result shows that the spin entanglement of these
states (for a given momentum $\vec{p}$) is not relativistic
invariant. By using the prescriptions of References \cite{reza},
\cite{reza1}, the introduced EWs can be manipulated via the LP
which can be solved exactly via simplex method. The region of
entangled states which can be detected via each kind of EWs is
determined. It is shown that, bipartite non-detectable density
matrices of the same type (their structures are the same except
for the positivity of density matrices) as the EWs of the first
kind are necessarily separable. Also, we discuss the
decomposability or non-decomposability of the EWs, where the
region of non-decomposable EWs of the first kind is partially
determined and it is shown that, all of the EWs of the second kind
are decomposable. It should be noticed that, without using the
techniques such as LP optimization method construction of optimal
EWs specially non-decomposable ones is not an easy task and as far
as we know, this work is a first step toward a relativistic
extension of quantum entanglement in multispinor systems with
mixed states specially PPT mixed ones. Moreover, similar to the
References \cite{reza}, \cite{reza1} and \cite{jsn}, one can
obtain some decomposable or non-decomposable positive maps from
the introduced multispinor EWs by using the Jamio\l kowski
isomorphism \cite{14}, \cite{5} but this is not treated in this
work. We discuss also examples of EWs (in each kind) for which the
feasible regions are not polygon and so, the region of EWs can be
approximately determined by LP (in these cases, the convex
optimization is reduced to the LP one). It is shown that, in
contrary to the exactly manipulatable EWs, both the first and
second kind of the optimal approximate EWs can detect some PPT
entangled states.

The paper is organized as follows: In section $2$, some of the
definitions and properties related to the EWs, linear programming
(LP) and general scheme for manipulation of EWs by using the exact
and approximate LP method are reviewed. In section $3$, two kinds
of BSD multispinor EWs in space-time with arbitrary dimension are
introduced. Also, the optimality of some of these EWs in each kind
is proved. Section $4$ is devoted to the region of entangled
states which can be detected by the introduced EWs. In particular,
in the bipartite systems in four-dimensional space-time, the
Bell-type and iso-concurrence type entangled states are defined
and it is shown that the amount of spin entanglement measured by
the Hilbert-Schmidt measure is not Lorentz invariant. In section
$5$, the decomposability or non-decomposability of the introduced
EWs is discussed. In section $6$ by using the approximate LP, two
new kinds of mutispinor EWs are manipulated. Section $7$ is
devoted to a brief discussion about systems with the odd number of
the spinors. The paper is ended with a brief conclusion and five
appendices.
\section{Preliminaries}
In this section, we briefly mention those concepts and subjects such
as definitions and properties related to the EWs and their
manipulation via the LP method as will be needed in the sequel; a
more detailed treatment may be found, for example, in
\cite{woro,20}.
\subsection{Multipartite systems and Entanglement Witnesses}
First we recall the notion of the separability for a system shared
by $N$ parties. Following Ref. \cite{29}, a $k$-partite split is a
partition of the system into $k\leq N$ sets $\{S_i\}_{i=1}^k$,
where each of them may be composed of several original parties.
Given a density operator $\rho_{1...k}\in
\mathcal{B}(H_1\otimes...\otimes H_k)$ the Hilbert space of
bounded operators acting on $H_1\otimes...\otimes H_k$ associated
with some $k$-partite split, we say that $\rho_{1...k}$ is a
$m$-separable state if it is possible to find a convex
decomposition for it such that in each pure state term at most $m$
parties are entangled among each other, but not with any member of
the other group of $N-m$ parties. For example, every $1$-separable
state, also called fully separable, can be written as
\begin{equation}\label{fullsep}
\rho_{1...k}=\sum_ip_i|\psi_i\rangle_1\langle\psi_i|\otimes...\otimes
|\psi_i\rangle_k\langle\psi_i|
\end{equation}
with $p_i\geq0$ and $\sum_ip_i=1$, hence, the set of all fully
separable states (hereafter, the separable states mean the fully
separable states) is a convex set called the convex set of separable
states ($CSSS$).\\
\textbf{Definition 1.} A Hermitian operator $W$ is called an EW
detecting the entangled state $\rho_{e}$ if \ $
\mathrm{Tr}(W\rho_{e})<0 $ and $ \mathrm{Tr}(W\rho_{s})\geq 0 $ for
all separable states $\rho_{s}$.

This definition has a clear geometrical meaning. The expectation
value of an observable depends linearly on the state. Thus, the
set of states for which $\mathrm{Tr}(W\rho)=0$ holds is a
hyperplane in the set of all states, cutting this set into two
parts. The part with $\mathrm{Tr}(W\rho)>0$ contains the set of
all separable states where the other part ( with
$\mathrm{Tr}(W\rho)<0$) is the set of states detectable by $W$.
From this geometrical interpretation it follows that for each
entangled state $\rho_{e}$, there exists an EW detecting it \cite{Horod2}.\\
\textbf{Definition 2.} An EW $W$ is decomposable (d-EW) iff there
exists operators $P,\ Q_{i}$ with
\begin{equation}\label{decom}
W=P+Q_{1}^{T_{A}}+Q_{2}^{T_{B}}+...+Q_{N}^{T_{Z}}
 \quad\quad P,Q_{i}\geq0
\end{equation}
where superscripts $T_i$ denote partial transposition with respect
to the subsystem $i$. $W$ is non-decomposable EW if it can not be
put in the form (\ref{decom}) (for more details see
\cite{Doherty3}).

One should notice that, only non-decomposable EWs can detect PPT
entangled states \cite{woro}. Then, an EW is nondecomposable
(nd-EW) iff there exists at least one PPT
entangled state which the witness detects \cite{woro}.\\
\textbf{Definition 3.} An EW $W$ is said to be optimal and denoted
by $W_{opt.}$ if for all positive operators $P$ and
$\varepsilon>0$, the following new Hermitian operator
\begin{equation}\label{Wn2}
W_{new}=(1+\varepsilon) W_{opt.}-\varepsilon P
\end{equation}
is not anymore an EW \cite{14}.

Suppose that there is a positive operator $P$ and $\epsilon\geq0$
such that  $W_{new}=\; W_{opt.}-\epsilon P $ is yet an EW. This
means that if $Tr(W_{opt.}\rho_s)=0$, then $Tr(P\rho_s)=0$, for
all separable states $\rho_s$. By using the fact that every
separable state is convex combination of pure product states, one
can take $\rho_s$ as a pure product state
$\ket{\psi}\langle\psi|$. Also, one can assume that the positive
operator $P$ is a pure projection operator, since an arbitrary
positive operator can be written as convex combination of pure
projection operators with positive coefficients.
\subsection{Manipulating EWs by exact and approximate LP method}Consider a
Hermitian operator $W$ with some negative eigenvalues as
\begin{equation}\label{HW}
 W=a_0I+\sum_{i=1}^na_{i} Q_{i}
\end{equation}
where $Q_{i}$ are Hermitian operators which will be considered as
multiplications of the Dirac $\gamma$ matrices,
 with $-1\leq \mathrm{Tr}( Q_{i} \rho_{s})\leq1$ for
each separable state $ \rho_{s}$ and $a_i$'s are real parameters
with $a_0\geq0$.

As $\rho_s$ varies over $CSSS$, the map $P_i=Tr(Q_i\rho_s)$ maps
$CSSS$ into a convex region called feasible region (inside the
hypercube defined by $-1\leq P_i\leq1$). Now, we try to choose the
real parameters $a_i$, $i=1,...,n$ (the allowed values of $a_i$
define a region called EW's region in the space of the parameters
$a_i$) such that the operator $W$ given in (\ref{HW}) possesses at
least one negative eigenvalue and its expectation value over any
separable state be non-negative, i.e., the condition
$Tr(W\rho_s)=a_0+\sum_{i=1}^na_iP_i\geq0$ be satisfied for all
$P_i$ belonging to the feasible region. The region of the
parameter space where, $W$ possesses non-negative expectation
value over all separable states (containing the EWs' region), is
called the region of separable states non-negative expectation
valued (denoted by SSNNEV).

Therefore, for determination of EWs of type (\ref{HW}), one needs to
determine the minimum value of $a_0+\sum_{i=1}^na_iP_i$ over the
feasible region (the minimum value must be non-negative) and hence
the problem reduces to the optimization of the linear function
$a_0+\sum_{i=1}^na_iP_i$ over the convex set of feasible region.

We note that, the minimum value of $F_W:=\mathrm{Tr}(W\rho_{s})$
achieves for pure product states, since every separable state
$\rho_{s}$ can be written as a convex combination of pure product
states (due to the convexity of separable region) as
$\rho_{s}=\sum_{i}p_{i} |\psi_{i}\rangle\langle\psi_{i}|$ with
$p_{i}\geq0$ and $\sum_{i}p_{i}=1$,  hence we have
\begin{equation}
F_W=\sum_{i}p_{i}\mathrm{Tr}(W|\psi_{i}\rangle\langle\psi_{i}|)\geq
C_{min}
\end{equation}
with $C_{min}=\min_{_{|\psi\rangle\in D_{prod.}}} \mathrm{Tr}(W
|\psi\rangle\langle\psi|)$, where $D_{prod.}$ denotes the set of
pure product states. Thus we need to find the pure product state
$|\psi_{min}\rangle$ which minimizes
$\mathrm{Tr}(W|\psi\rangle\langle\psi|)$. For the cases that the
feasible regions are simplexes (or at most convex polygons), the
manipulation of the EWs amounts to
$$
\hspace{-1.75cm}\mathrm{minimize}\quad\quad
F_{_{W}}=a_0+\sum_{i=1}^na_{i}P_{i}$$
\begin{equation}\label{LP}
\mathrm{subject \ to}\quad \sum_{i=1}^n(c_{ij}P_{i}-d_{i})\geq 0
\quad  j=1,2,...
\end{equation}
where $c_{ij}$ and $d_{i}$, $i,j=1,2,...$ are parameters of
hyperplanes surrounding the feasible regions.

One can calculate the distributions $P_i$, consistent with the
aforementioned optimization problem, from the information about
the boundary of feasible region. To achieve the feasible region we
obtain the extreme points corresponding to the product
distributions $P_i$ for every given product state by applying the
special conditions on the parameters $a_i$. In fact, $F_{_W}$
themselves are functions of the product distributions, and they
are in turn functions of $\psi$. They are not real variables of
$\psi$ but the product states will be multiplicative. If this
feasible region constructs a polygon by itself, the corresponding
boundary points of the convex hull will minimize exactly $F_{_W}$
in Eq.(\ref{LP}). This problem is called exact LP and the simplex
method is the easiest way of solving it \cite{Boyd}.

If the feasible region is not a polygon, with the help of tangent
planes in this region at points which are determined either
analytically or numerically, one can define a new convex hull
which is a polygon encircling the feasible region. The points on
the boundary of the polygon can approximately determine the
minimum value of $F_W$ in (\ref{LP}). Thus the approximated value
is obtained via LP.
\section{Entanglement witnesses for relativistic and non-relativistic multispinor systems in space-time dimension $d$}
In this section, first we introduce our general formalism for
constructing multispinor EWs by using Dirac $\gamma$ matrices. In
general we consider $m$ spinors in the space-time of dimension $d$
and $D^m$ dimensional Hilbert space
$\mathcal{H}=\underbrace{\mathcal{H}_D\otimes...\otimes
\mathcal{H}_D}_{m}$ with $D=2^{\lfloor d/2\rfloor}$.

Let $\gamma_{\mu}$, $\mu=1,...,d$, be $d$ Dirac $\gamma$ matrices
satisfying the anticommuting relations:
\begin{equation}\label{eqqq00}
\gamma_{\mu}\gamma_{\nu}+\gamma_{\nu}\gamma_{\mu}=2\delta_{\mu\nu}I.
\end{equation}
It follows from relations (\ref{eqqq00}) that the $\gamma$ matrices
$\gamma_{\mu}$ generate an algebra which, as a vector space, has a
dimension $2^{d}$ (for a brief review on the Dirac $\gamma$ matrices
see appendix $A$). We consider hermitian matrices $A_i$, $i=1,2,...,
2^{d}$ as all possible multiplications of $\gamma_{\mu}$,
$\mu=1,2,...,d$ up to multiplicative factors $\pm1,\pm i$. Then, we
will have
\begin{equation}\label{gama'}
A^{2}_i=I_{2^{\lfloor d/2\rfloor}},\;\;\ A_i=A^{\dag}_i,\;\;\
tr(A_iA_j)=2^{\lfloor d/2\rfloor}\delta_{ij}\;\ \mbox{for all} \;\
i,j=1,2,...,2^{d}.
\end{equation}
Clearly, the operators $A_i$, $i=1,2,...,2^d$ either commute or
anti-commute with each other, hence for even number of spinors,
the matrices $\underbrace{A_i\otimes...\otimes A_i}_m$, $i=1,...,
2^{d}$ commute with each other and can be diagonalized
simultaneously. Also note that, we have $(A_i\otimes...\otimes
A_i)^2=I_{2^{m\lfloor d/2\rfloor}}$, therefore the eigenvalues of
$A_i\otimes...\otimes A_i$, $i=1,...,2^{d}$ are $\pm 1$. In order
to construct Bell-states diagonal multispinor EWs, we will
consider a hermitian operator $W$ as superposition of the
operators $A_i\otimes...\otimes A_i$, $i=1,...,2^{d}$ such that
the conditions of definition $1$ are satisfied. It will be seen
that for some suitable superpositions of the operators
$A_i\otimes...\otimes A_i$, the manipulation of the EWs reduces to
the linear programming which can be solved exactly by using the
simplex method.

It should be noticed that the linear combination of product of
locally commuting matrices $A_i$, i.e.,
\begin{equation}\label{op0}
W=\sum_ia_i A^{(1)}_i\otimes ...\otimes A^{(m)}_i
\end{equation}
with $[A^{(k)}_i,A^{(k)}_j]=0$ for $i\neq j$, $k=1,...,m$ (the
upper index $(k)$ denotes the $k$-th spinor), can not be an EW,
since its eigenvalues are all positive. In fact, $W$ in
(\ref{op0}) can be written as
\begin{equation}\label{op1}
W=\sum_{i;\alpha_1,...,\alpha_m}a_i
\lambda_{i\alpha_1}...\lambda_{i\alpha_m}
E^{(1)}_{\alpha_1}\otimes...\otimes E^{(m)}_{\alpha_m},
\end{equation}
where, $E^{(k)}_{\alpha_i}$, $k=1,...,m$ is the projection
operator to the eigenspace of $A^{(k)}_i$ corresponding to the
eigenvalue $\lambda_{i\alpha_k}$. Then, the non-negativity of
$\mathrm{Tr}(W|\alpha_1\rangle\langle\alpha_1|\otimes...\otimes
|\alpha_m\rangle\langle\alpha_m|)\geq0$ implies that $\sum_{i}a_i
\lambda_{i\alpha_1}...\lambda_{i\alpha_m}\geq0$, i.e., the
eigenvalues of $W$ are all positive and so $W$ can not be an EW.
\subsection{Two particular sets of operators} In the following we choose two
particular sets of the above introduced operators $A_i$,
$i=1,2,...,2^{d}$ which will be used in manipulating the
multispinor EWs via exact LP optimization method.
\subsubsection{First kind: Maximally anticommuting sets}
As it was mentioned before, in the Hilbert space of dimension
$D=2^{\lfloor d/2\rfloor}$, we have $d$ matrices $\gamma_{\mu}$,
$\mu=1,...,d$ which anticommute with each other. In the case of
even dimension $d$, we denote
$\gamma_S:=i^{-d/2}\gamma_1\gamma_2...\gamma_{d}$ by
$\gamma_{d+1}$, then the matrices $A_i=\gamma_i$, for
$i=1,2,...,d,d+1$ form a maximally anticommuting set in the
algebra of $\gamma$ matrices (in the case of odd $d$, the set of
matrices $\gamma_i$, $i=1,...,d$ is maximally anticommuting set).

It is well known that every solution for the anticommutation
relations (\ref{eqqq00}) is equivalent to one another. That is if
$\gamma_{\mu}$ and $\gamma'_{\mu}$ be two solutions for
(\ref{eqqq00}), then there exists a unitary matrix $U$ such that
\begin{equation}\label{eqqq'}
\gamma'_{\mu}=U\gamma_{\mu}U^{-1},\;\;\ (\mu=1,2,...,d).
\end{equation}
For proof see Ref. \cite{ohnuki}. Therefore, every EW defined as a
superposition of the matrices $\gamma_{\mu}$, can be replaced with
another equivalent one in which the matrices $\gamma_{\mu}$ are
replaced with the matrices $\gamma'_{\mu}$. Also, we will use the
fact that, for any two anticommuting hermitian operators $A$ and
$B$, the expectation value of $B$ over any eigenvector $\ket{\psi}$
of $A$ with eigenvalue $\lambda$ is zero and vice versa. Explicitly,
we have
\begin{equation}\label{P2'}
0=\langle\psi|(AB+BA)|\psi\rangle=2\lambda\langle\psi|B|\psi\rangle.
\end{equation}
\subsubsection{Second kind: Commuting sets which
anticommute with each other} The second kind of sets for which the
EW can be manipulated via exact LP, is the sets which are the
union of three commuting sets $C_1,C_2, C_3$ such that
$\{C_i,C_j\}=0$ for $i\neq j$, i.e.,  for each $x,y\in C_i$,
$i=1,2,3$  we have $[x,y]=0$, while for each $x\in C_i$, $z\in
C_j$, $j\neq i$ we have $\{x,z\}=0$.

Clearly, for a given $d$, there are several such commuting sets.
In this paper, we will consider the following commuting sets
$C_i$, $i=1,2,3$ for constructing the multispinor EWs of the
second kind:
$$C_1=\{-i\gamma_{1} \gamma_{2}, -\gamma_{1} \gamma_{2} \gamma_{3} \gamma_{4},...,i^{-\lfloor d/2\rfloor}\gamma_{1}
\gamma_{2}...\gamma_{2\lfloor d/2\rfloor}\},\;\;\
C_2=\{\gamma_{1}, i\gamma_{1} \gamma_{3}
\gamma_{4},...,i^{-\lfloor d/2\rfloor-1}\gamma_{1}
\gamma_{3}...\gamma_{2\lfloor d/2\rfloor}\},$$
\begin{equation}\label{gamac}
C_3=\{\gamma_{2}, -i\gamma_{2} \gamma_{3}
\gamma_{4},...,i^{-\lfloor d/2\rfloor+1}\gamma_{2}
\gamma_{3}...\gamma_{2\lfloor d/2\rfloor}\}.
\end{equation}
Note that each set $C_i$, $i=1,2,3$ has cardinality $\lfloor
d/2\rfloor$.
\subsection{Construction of BSD multispinor EWs}
In this section we consider $m$ $d$-dimensional spinors in the
Hilbert space of dimension $2^{m\lfloor d/2\rfloor}$ and construct
EWs by using the two sets of hermitian operators introduced in the
previous subsection. In the following, we will consider only the
case of even $m$ in details, where all of the discussions can be
applied in the case of odd $m$, straightforwardly. In section $7$,
we will discuss the case of odd $m$ briefly. Also, in the rest of
the paper, we will consider even space-time dimensions $d$ in
order to simplify the notations. All of discussions and the
equations given for even $d$ such as the form of the introduced
EWs, density matrices, etc. are the same for odd dimensions only
by replacing $d$ with $d-1$.
\subsubsection{EWs of the first kind}
In the case of even $m$, we will consider the following hermitian
matrix
\begin{equation}\label{Wn}
W^{(m;d)}=a_0 I_{2^{m d/2}}+\sum_{i=1}^{d+1} a_i
\underbrace{\gamma^{(d)}_i\otimes \gamma^{(d)}_i\otimes...\otimes
\gamma^{(d)}_i}_m,
\end{equation}
where, $\gamma^{(d)}_i$ for $i=1,..,d+1$ are Dirac $\gamma$
matrices in the space-time of even dimension $d$. In order that
the observable (\ref{Wn}) turns to an EW, we need to choose its
parameters in such a way that it becomes a non-positive operator
with non-negative expectation values over any separable state
$\rho_s$.
\\
Now it is the time to reduce the problem to the LP one. In order
to determine the feasible region,  we need to know the apexes,
namely the extreme points, to construct the hyperplanes
surrounding the feasible region.

For a given separable state $\rho_s$, the non-negativity of
\begin{equation}\label{W2''}
\mathrm{Tr}(W^{(m;d)}\rho_s)\geq 0,
\end{equation}
implies that
\begin{equation}\label{W3}
a_0+\sum_{i=1}^{d+1}a_iP_i\geq0,
\end{equation}
with
\begin{equation}\label{P2}
P_i=tr(\rho_s\gamma_i^{(d)}\otimes ...\otimes \gamma_i^{(d)}),
\end{equation}
where all of the $P_i$'s lie in the interval $[-1,1]$ (since, the
eigenvalues of $\gamma_i^{(d)}\otimes ...\otimes \gamma_i^{(d)}$
are $\pm 1$). Now, by using the fact that $\gamma_i^{(d)}$'s
anticommute with each other and therefore the expectation value of
$\gamma_j^{(d)}$ over any eigenvector of $\gamma_i^{(d)}$, $i\neq
j$ is zero, one can deduce that the extremum points or apexes are
given as follows
\begin{equation}\label{Tab1}
\begin{tabular}{c|c}\hline\hline
  Product state & $(P_1,P_2,...,P_{d+1})$ \\ \hline
  $|\psi^{(1)}_{\pm}\rangle$ & $(\pm1,0,0,0,...,0,0,0) $ \\
  $|\psi^{(2)}_{\pm}\rangle$ & $(0,\pm1,0,0,...,0,0,0)$ \\
  $\vdots$ & $\vdots$ \\
  $|\psi^{(k)}_{\pm}\rangle$ & $(0,...,0,\underbrace{\pm1}_{k-th},0,...,0)$ \\
  $\vdots$ & $\vdots$\\
  $|\psi^{(d+1)}_{\pm}\rangle$ & $(0,0,0,...,0,0,0,\pm1)$\\
         \hline\hline
\end{tabular}
\end{equation}
where, $\ket{\psi^{(i)}_{\pm}}$ are eigenvectors of
$\gamma^{(d)}_i\otimes...\otimes \gamma^{(d)}_i$ with eigenvalues
$\pm 1$.

Regarding the above consideration, we are now ready to state the
feasible region which is the convex hull of the apexes given in
(\ref{Tab1}).  According to the following inequalities
\begin{equation}\label{ine0'}
Tr\{\rho_s(I+\sum_{k=1}^{d+1}(-1)^{i_{k}}\underbrace{\gamma^{(d)}_i\otimes
...\otimes
\gamma^{(d)}_i}_m)\}=1+\sum_{k=1}^{d+1}(-1)^{i_{k}}P_{_k}\geq
0,\quad \forall \ (i_{1},i_{2},...,i_{d},i_{d+1})\in
\{0,1\}^{d+1},
\end{equation}
(for the proof, see appendix $B$) any separable state is mapped
into halfspaces defined by $
1+\sum_{k=1}^{d+1}(-1)^{i_{k}}P_{_k}\geq 0 $ and consequently, the
feasible region corresponds to the intersection of these
halfspaces which is the convex hull of the apexes. Therefore, the
feasible region is surrounded by $2^{d+1}$ hyperplanes defined in
a space of dimension $d+1$ as follows
\begin{equation}\label{ine02}
1+\sum_{k=1}^{d+1}(-1)^{i_{k}}P_{_k}=0 \quad,\quad \forall \
(i_{1},i_{2},...,i_{d},i_{d+1})\in \{0,1\}^{d+1}.
\end{equation}

Now, according to the prescription of subsection $2.2$, namely the
equation (\ref{LP}), the non-negativity of $W^{(m;d)}$ over
separable states can be achieved by solving the following LP problem
$$
\hspace{-6cm}\mathrm{minimize} \quad\quad\;\
a_0+\sum_{i=1}^{d+1}a_iP_i \vspace{-3mm}
$$
\begin{equation}\label{region0}
 \hspace{-2cm}\mathrm{subject\ to} \quad\;\;\left\{\begin{array}{c}
 \hspace{-1.5cm}1+\sum_{k=1}^{d+1}(-1)^{i_k}P_k\geq 0\\
 \forall\ |P_k|\leq 1, \;\; k=1,...,d,d+1 \hspace{1cm}\\
\end{array}\right.
\end{equation}
with $i_k\in\{0,1\}$.

In the appendix $C$ it is shown that any vertex point of the
feasible region corresponds to a hyperplane of the region of
SSNNEV and each hyperplane corresponding to the feasible region
(e.g., each of the $2^{d+1}$ hyperplanes given in (\ref{ine02}))
corresponds to an extreme point of this region. Therefore, by
substitution of vertex points of the feasible region given in
(\ref{Tab1}), we get the region of SSNNEV as the intersection of
the following halfspaces
\begin{equation}\label{a2}
|a_{_i}|\leq a_0,\quad i=1,2,...,d+1.
\end{equation}
The above inequalities imply that in the space of parameters $a_i$
of EWs, by fixing $a_{0}$, all of the other $a_i$'s lie inside the
hypercube $|a_{i}|\leq a_0$, $i=1,...,d+1$. Also, we will need all
eigenvalues of $W^{(m;d)}$ which consist of
\begin{equation}\label{Wn2xxx}
\lambda^{(m;d)}_{i_1...i_{d}}=a_0 +\sum_{k=1}^{d}(-1)^{i_k}
a_k+i^{-m d/2}(-1)^{i_1+i_2+...+i_{d}}a_{d+1},
\end{equation}
where $i_k\in \{0,1\}$, $k=1,2,...,d$. Therefore, at least one of
the eigenvalues $\lambda^{(m;d)}_{i_1...i_{d}}$ must be negative
to be guarantied $W^{(m;d)}$ is an EW. We note that, the
intersection of $2^{d}$ halfspaces defined by
$\lambda^{(m;d)}_{i_1...i_{d}}\geq 0$ is the region of
$W^{(m;d)}\geq 0$ which is a polytope. Then, the complement of
this polytope in the $d+1$ dimensional hypercube defined by
$|a_i|\leq a_0$, $i=1,2,...,d+1$ is the region of EWs (clearly,
the region of EWs is nonempty since $2^{d}< 2^{d+1}$). Also, it
can be noticed that the optimal EWs are the farthest ones from the
region $W^{(m;d)}\geq 0$, i.e., the vertex points of the the EWs'
region.

Moreover, Eq.(\ref{ine0'}) shows that the region of SSNNEV
(hypercube) has $2^{d+1}$ extreme points as
$((-1)^{i_1},...,(-1)^{i_{d}},(-1)^{i_{d+1}})$ with
$i_1,...,i_{d+1}\in \{0,1\}$. In fact, the half of these points
corresponds to the positive operators, where the other half of
them corresponds to the extreme points of the EW's region, i.e.,
optimal EWs. These $2^d$ extreme points are given by
$((-1)^{i_1},(-1)^{i_2},...,(-1)^{i_{d}},-(-i)^{md/2}(-1)^{i_1+...+i_{d}})$
with $i_1,...,i_{d}\in \{0,1\}$ corresponding to the following
$2^{d}$ optimal EWs:
\begin{equation}\label{region}
W^{(m;d;i_1,...,i_{d})}_{opt.}=I_{2^{m
d/2}}+\sum_{k=1}^{d}(-1)^{i_k}\underbrace{\gamma^{(d)}_k\otimes...\otimes
\gamma^{(d)}_k}_m-(-i)^{m
d/2}(-1)^{i_1+...+i_{d}}\underbrace{\gamma^{(d)}_{d+1}\otimes...\otimes
\gamma^{(d)}_{d+1}}_m,
\end{equation}
where, $i_1,...,i_{d}\in \{0,1\}$. We will prove the optimality of
$W^{(m;d;i_1,...,i_{d})}_{opt.}$ in subsection $3.3$.
\subsubsection{EWs of the second kind} Now, we consider a superposition of the second set
of operators introduced in subsection $3.1.2$ as follows
\begin{equation}\label{EW'}
W'^{(m;d)}=a'_0I_{2^{md/2}}+\sum_{i=1}^{3
d/2}a'_i\underbrace{A'_i\otimes A'_i\otimes...\otimes A'_i}_{m},
\end{equation}
where, $A'_1,...,A'_{d/2}\in C_1$, $A'_{d/2+1},...,A'_{d}\in C_2$
and $A'_{d+1},...,A'_{3d/2}\in C_3$. Note that these $3 d/2$
operators do not form independent generating set, namely, we have
\begin{equation}\label{eqx}
A'_{d/2+i}=(-1)^{i-1}A'_{d/2+1}A'_iA'_1,\;\;\ A'_{d+i}=iA'_{
d/2+1}A'_i\;\ \mbox{for}\;\ i=1,2,...,d/2.
\end{equation}

In order that $W'^{(m;d)}$ be an EW, the expectation value of it
on any separable state must be non-negative, i.e., for a given
separable state $\rho_s$, the condition
\begin{equation}\label{W3'}
a'_0+\sum_{i=1}^{3d/2}a'_iP'_i\geq0,
\end{equation}
must be hold where, $ P'_i:=tr(\rho_sA'_i\otimes ...\otimes
A'_i)$. Clearly we have $|P'_i|\leq 1$ for $i=1,...,3d/2$, since
the eigenvalues of $A'_i\otimes ...\otimes A'_i$ are $\pm1$.

The extremum points or apexes are given by
\begin{equation}\label{tab200}
\begin{tabular}{c|c}\hline\hline
  Product state & $(P'_1,...P'_{d/2};P'_{d/2+1},...,P'_{d};P'_{d+1},...,P'_{3d/2})$ \\ \hline
  $|\psi^{(1;1)}_{\pm}\rangle$ & $(\pm 1,1,1,...,1; 0,0,...,0; 0,0,...,0)$ \\
  $\vdots$ & $\vdots$ \\
  $|\psi^{(1;d/2)}_{\pm}\rangle$ & $(1,...,1,1,\pm 1;0,0,...,0;0,...,0,0)$ \\
  $|\psi^{(2;d/2+1)}_{\pm}\rangle$ & $(0,0,...,0;\pm 1,1,1,...,1;0,0,...,0)$\\
  $\vdots$ & $\vdots$\\
  $|\psi^{(2;d)}_{\pm}\rangle$ & $(0,0,...,0;1,1,...,1,\pm 1;0,0,...,0)$\\
  $|\psi^{(3;d+1)}_{\pm}\rangle$ & $(0,0,...,0;0,...,0,0;\pm 1,1,...,1,1)$\\
  $\vdots$ & $\vdots$\\
  $|\psi^{(3;3d/2)}_{\pm}\rangle$ & $(0,0,...,0;0,0,...,0;1,...,1,1,\pm 1)$\\
         \hline\hline
\end{tabular},
\end{equation}
where, $\ket{\psi^{(i;k)}_{\pm}}$ for $i=1,2,3$;
$k=1+(i-1)d/2,..., id/2$ are common eigenvectors of the elements
of the commuting set $C_i$ such that
\begin{equation}
A'_j\ket{\psi^{(i;k)}_{\pm}}=(\pm
1)^{\delta_{jk}}\ket{\psi^{(i;k)}_{\pm}},\;\;\ A'_j\in C_i.
\end{equation}
Note that, we have used the fact that the elements of $C_i$
commute with each other and anticommute with the elements of
$C_j$, for $j\neq i$.

Considering the apexes given by (\ref{tab200}), one can obtain the
following inequalities
$$Tr\{\rho_s(I+(-1)^{i_1} A'_j\otimes...\otimes
A'_j+(-1)^{i_2}A'_{j+d/2}\otimes...\otimes A'_{j+d/2}+(-1)^{i_3}
A'_{j+d}\otimes...\otimes A'_{j+d})\}=$$
\begin{equation}\label{Cn}
1+(-1)^{i_1} P'_j+(-1)^{i_2}P'_{j+d/2}+(-1)^{i_3} P'_{j+d}\geq0,
\end{equation}
where, $i_1,i_2,i_3\in\{0,1\}$ and $j\in \{1,...,d/2\}$ (for the
proof, see appendix $B$). Therefore, the feasible region is the
intersection of the halfspaces given by (\ref{Cn}) and the
hyperplanes surrounding the feasible region are as follows
\begin{equation}\label{Cn'}
1+(-1)^{i_1} P'_j+(-1)^{i_2}P'_{j+d/2}+(-1)^{i_3} P'_{j+d}=0.
\end{equation}

Again, in order to manipulate the EWs, according to the equation
(\ref{LP}) one needs to solve the following LP problem
$$
\hspace{-8cm}\mathrm{minimize} \quad\;\;\;\;\;\
a'_0+\sum_{i=1}^{3d/2}a'_iP'_i \vspace{-3mm}
$$
\begin{equation}\label{region1}
\hspace{2cm} \mathrm{subject\ to} \quad\;\;\left\{\begin{array}{c}
1+(-1)^{i_1}P'_j+(-1)^{i_2}P'_{j+d/2}+(-1)^{i_3} P'_{j+d}\geq 0\\
 \hspace{-6.5cm}\forall\ |P'_k|\leq 1,\\
\end{array}\right.
\end{equation}
with $i_1,i_2,i_3\in\{0,1\}$ and $j\in \{1,...,d/2\}$.

Putting the coordinates of the apexes of the feasible region given
by (\ref{tab200}) in Eq.(\ref{W3'}), yields the region of SSNNEV
as the intersection of the following halfspaces
\begin{equation}\label{vp2'}
|\sum_{k=1}^{d/2}(-1)^{i_k}a'_k|\leq a'_0,\quad
|\sum_{k=1}^{d/2}(-1)^{i_k}a'_{d/2+k}|\leq a'_0,\quad
|\sum_{k=1}^{d/2}(-1)^{i_k}a'_{d+k}|\leq a'_0.
\end{equation}
We will also need all of the eigenvalues of $W'^{(m;d)}$ which
consist of
\begin{equation}\label{Wn2xx}
\lambda^{'(m;d)}_{i_1...i_{d/2+1}}=a'_0
+\sum_{k=1}^{d/2+1}(-1)^{i_k}
a'_k+\sum_{k=2}^{d/2}(-1)^{i_1+i_{d/2+1}+i_{k}}a'_{d/2+k}+\sum_{k=1}^{d/2}(-1)^{m/2+i_{d/2+1}+i_{k}}a'_{d+k},
\end{equation}
where $i_1,...,i_{d/2+1}\in \{0,1\}$ (we have used the
Eq.(\ref{eqx})). Again, in order that $W'^{(m;d)}$ be an EW, at
least one of the eigenvalues $\lambda^{'(m;d)}_{i_1...i_{d/2+1}}$
must be negative. In fact, the intersection of $2^{d/2+1}$
halfspaces defined by $\lambda^{'(m;d)}_{i_1...i_{ d/2+1}}\geq 0$
is the region of $W'^{(m;d)}\geq 0$ which is a polytope. Then, the
complement of this polytope in the region defined by (\ref{vp2'})
is the region of EWs.

Also, the inequalities (\ref{Cn}) imply that the region of SSNNEV
has $8.d/2=4d$
vertices as\\
$(0,...,0,\underbrace{(-1)^{i_1}}_j,0,...,0;0,...,0,\underbrace{(-1)^{i_2}}_{j+d/2},0,...,0;0,...,0,\underbrace{(-1)^{i_3}}_{j+d},0,...,0)$
with $j\in \{1,...,d/2\}$ and $i_1,i_2,i_3\in \{0,1\}$. It can be
shown that, half of these points corresponds to the positive
operators, where the other half corresponds to the optimal EWs. In
fact we have $2d$ extreme points as
$(0,...,0,\underbrace{(-1)^{i_1}}_j,0,...,0;0,...,0,\underbrace{(-1)^{i_2}}_{j+d/2},0,...,0;0,...,0,\underbrace{-(-1)^{m/2+i_1+i_2}}_{j+d},0,...,0)$
with $j\in \{1,...,d/2\}$ and $i_1,i_2\in \{0,1\}$ corresponding
to the following $2d$ optimal EWs:
\begin{equation}\label{W'n}
W'^{(m;d;i_1,i_2;j)}_{opt.}=I+(-1)^{i_1}\underbrace{A'_j\otimes...\otimes
A'_j}_m+(-1)^{i_2}\underbrace{A'_{j+d/2}\otimes...\otimes A'_{j+
d/2}}_m-(-1)^{m/2+i_1+i_2}\underbrace{A'_{j+d}\otimes...\otimes
A'_{j+d}}_m,
\end{equation}
with $i_1,i_2\in \{0,1\}$, $j\in \{1,...,d/2\}$. We prove the
optimality of these EWs in the following subsection.
\subsection{Optimality of EWs $W^{(m;d;i_1,...,i_d)}_{opt.}$ and $W'^{(m;d;i_1,i_2;j)}_{opt.}$}
In this section, we prove the optimality of EWs
$W^{(m;d;i_1,...,i_d)}_{opt.}$ and $W'^{(m;d;i_1,i_2;j)}_{opt.}$
given by (\ref{region}) and (\ref{W'n}), respectively.
\subsubsection{Optimality of $W^{(m;d;i_1,...,i_d)}_{opt.}$}
In order to prove that the $W^{(m;d;i_1,...,i_d)}_{opt.}$ given in
Eq.(\ref{region}) is optimal, we first rewrite
$W^{(m;d;i_1,...,i_d)}_{opt.}$ as follows
\begin{equation}\label{Wn2x}
W^{(m;d;i_1,...,i_d)}_{opt.}=I+\sum_{k=1}^{d}(-1)^{i_k}O_k-(-i)^{md/2}(-1)^{i_1+...+i_{d}}O_{d+1},
\end{equation}
where, $O_i:=\underbrace{\gamma^{(d)}_i\otimes...\otimes
\gamma^{(d)}_i}_m$ for $i=1,2,...,d,d+1$ and prove the optimality
of
\begin{equation}\label{Wop1x}
W^{(m;d;1,...,1)}_{opt.}=I-\sum_{k=1}^{d}O_k-(-i)^{md/2}O_{d+1},
\end{equation}
where the optimality of the other cases can be proved similarly.
According to the definition 3 of subsection $2.1$, it suffices to
show that there exists no positive operator $P$ such that
$W_{new}:=(1+\varepsilon)W^{(m;d;1,...,1)}_{opt.}-\varepsilon P$
be an EW, namely it must be proved that for any pure product state
$|\psi\rangle$ such that
$Tr(W^{(m;d;1,...,1)}_{opt.}|\psi\rangle\langle\psi|)=0$, there
exists no positive operator $P$ with the constraint
$Tr(P|\psi\rangle\langle\psi|)=0$. To this end, first we note that
the expectation value of the operator $W^{(m;d;1,...,1)}_{opt.}$
in (\ref{Wop1x}) over pure product states $|\psi\rangle$ will
vanish if one of the equations
$$
O_i\ket{\psi}=\ket{\psi}\;\ \mbox{for some}\;\ i=1,2,...,d\;\
\mbox{or}
$$
\begin{equation}\label{2}
O_{d+1}\ket{\psi}=(-i)^{md/2}\ket{\psi}
\end{equation}
be satisfied (recall that $\langle\psi|O_j|\psi\rangle=0$, for
$j\neq i$, since $\ket{\psi}$ is a product state). Regarding the
definition $3$ of subsection $2.1$, we may assume that the
positive operator $P$ is a pure projection operator, since any
arbitrary positive operator can be written as convex combination
of pure projection operators with positive coefficients. The
equations (\ref{Wop1x}) and (\ref{2}) indicate that, in order that
$Tr(P\ket{\psi}\langle\psi|)=0$ be satisfied, the operator $P$
must be the projection operator to the eigenspace of $O_i$,
$i=1,2,...,d$ with eigenvalue $-1$ and $O_{d+1}$ with eigenvalue
$-(-i)^{md/2}$. But from the fact that
$O_{d+1}=(-i)^{md/2}O_{1}...O_{d}$, if $\ket{\psi'}$ be the common
eigenket of the operators $O_{1}, O_2,...,O_{d}$ with eigenvalue
$-1$, then $\ket{\psi'}$ will be an eigenket of $O_{d+1}$ with
eigenvalue $(-i)^{md/2}$
($O_{d+1}\ket{\psi'}=(-i)^{md/2}(-1)^{d}\ket{\psi'}=(-i)^{md/2}\ket{\psi'}$).
Therefore, the eigenspace of $O_i$, $i=1,2,...,d$ with eigenvalue
$-1$ and $O_{d+1}$ with eigenvalue $-(-i)^{md/2}$ is a null space.
\subsubsection{Optimality of $W'^{(m;d;i_1,i_2;j)}_{opt.}$}
We prove the optimality of $W'^{(m;d;i_1,i_2;j)}_{opt.}$ for $j=1$
and $i_1=i_2=1$, the optimality of the other cases can be proved
similarly. For $j=1$ and $i_1=i_2=1$, we have
\begin{equation}\label{Wop1}
W'^{(m;d;1,1;1)}_{opt.}=I-A'_1\otimes...\otimes
A'_1-A'_{d/2+1}\otimes...\otimes
A'_{d/2+1}-(-1)^{m/2}A'_{d+1}\otimes...\otimes A'_{d+1}.
\end{equation}
As regards the arguments of subsection $3.3.1$, we need to show
that the eigenspace of $A'_1\otimes...\otimes A'_1$,
$A'_{d/2+1}\otimes...\otimes A'_{d/2+1}$ with eigenvalue $-1$ and
$A'_{d+1}\otimes ...\otimes A'_{d+1}$ with eigenvalue
$-(-1)^{m/2}$ is a null space. Assume that $\ket{\psi'}$ be the
eigenket of $A'_1\otimes...\otimes A'_1$ and
$A'_{d/2+1}\otimes...\otimes A'_{d/2+1}$ with eigenvalue $-1$,
then by using (\ref{eqx}) we have
\begin{equation}\label{Wop1}
A'_{d+1}\otimes...\otimes
A'_{d+1}\ket{\psi'}=i^mA'_{d/2+1}A'_1\otimes...\otimes
A'_{d/2+1}A'_1\ket{\psi'}=(-1)^{m/2}\ket{\psi'}.
\end{equation}
This implies that, every eigenstate of $A'_1\otimes...\otimes
A'_1$ and $A'_{d/2+1}\otimes...\otimes A'_{d/2+1}$ with eigenvalue
$-1$ is necessarily an eigenstate of $A'_{d+1}\otimes...\otimes
A'_{d+1}$ with eigenvalue $(-1)^{m/2}$ and so the corresponding
common eigenspace is a null space.
\section{Entangled states which can be detected by BSD multispinor EWs}
In this section, we discuss the Bell-states diagonal entangled
states which can be detected by the introduced EWs. To do so,
first we consider the most significant case of bipartite system of
spinors in four-dimensional space-time and then generalize the
discussions to multipartite higher dimensional cases. In the
bipartite case, we use the Weyl or chiral representation of the
gamma matrices and follow the notation of the text by Weinberg
\cite{Weinberg} to take the Lorentz transformation of states more
conveniently. In the case of EWs of the first kind with $m=2,d=4$,
we consider both the relativistic and non-relativistic BSD density
matrices in order to discuss the effect of the Lorentz
transformation on the amount of entanglement measured by the
Hilbert-Schmidt measure, where for the case of EWs of the second
kind with $d=4$ and EWs with $d>4$, we discuss only the
non-relativistic density matrices which can be detected by these
EWs (and do not deal with the amount of entanglement), where the
discussions about relativistic case in $d=4$ can be generalized
straightforwardly to the cases $d>4$.
\subsection{Entanglement properties of relativistic and non-relativistic BSD density matrices in four-dimensional space-time}
In order to define some interesting entangled states detectable by
the introduced EWs, we construct Bell-type and iso-concurrence
type entangled states and investigate their entanglement
properties by using the introduced EWs (entanglement properties of
non-relativistic Bell-diagonal states and iso-concurrence states
have been studied in \cite{akhtar}- \cite{rez}). To this end, we
will take the helicity basis (simultaneously eigenstates of the
helicity operator \cite{ryder} and $\gamma'_5=(H\otimes
I)\gamma_5(H\otimes I)$, with
$H=\frac{1}{\sqrt{2}}(\sigma_x+\sigma_z)$ known as Hadamard
transform) and construct Bell-type and iso-concurrence type
entangled states by considering their combinations.

It is well known that, the helicity eigenstates \cite{ryder} are
given by
\begin{equation}\label{helicity}
\ket{\psi_1}=\frac{1}{\sqrt{2}}\left(\begin{array}{c}
                 1 \\
                 0 \\
                 1 \\
                 0 \\
               \end{array}\right),\quad \ket{\psi_2}=\frac{1}{\sqrt{2}}\left(\begin{array}{c}
                 0 \\
                 1 \\
                 0 \\
                 -1 \\
               \end{array}\right),\quad \ket{\psi_3}=\frac{1}{\sqrt{2}}\left(\begin{array}{c}
                 1 \\
                 0 \\
                 -1 \\
                 0 \\
               \end{array}\right),\quad \ket{\psi_4}=\frac{1}{\sqrt{2}}\left(\begin{array}{c}
                 0 \\
                 1 \\
                 0 \\
                 1 \\
               \end{array}\right)
\end{equation}
the first two of which correspond to positive energy, and the
second two to negative energy. One could notice that, the helicity
eigenstates $\ket{\psi_2},\ket{\psi_3}$ and $\ket{\psi_4}$ can be
obtained from $\ket{\psi_1}$ by local unitary transformations as
follows
\begin{equation}\label{bellloc}
\ket{\psi_2}=(\sigma_z\otimes \sigma_x)\ket{\psi_1},\quad
\ket{\psi_3}=(\sigma_z\otimes I)\ket{\psi_1},\quad
\ket{\psi_4}=(I\otimes\sigma_x)\ket{\psi_1}.
\end{equation} Now, we define the following Bell
states:
$$\ket{\psi_{\pm}}^{(1,2)}=\frac{1}{\sqrt{2}}(|\psi_1\rangle|\psi_1\rangle\pm|\psi_2\rangle|\psi_2\rangle),\quad
\ket{\phi_{\pm}}^{(1,2)}=\frac{1}{\sqrt{2}}(|\psi_1\rangle|\psi_2\rangle\pm|\psi_2\rangle|\psi_1\rangle),$$
$$\ket{\psi_{\pm}}^{(3,4)}=\frac{1}{\sqrt{2}}(|\psi_3\rangle|\psi_3\rangle\pm|\psi_4\rangle|\psi_4\rangle),\quad
\ket{\phi_{\pm}}^{(3,4)}=\frac{1}{\sqrt{2}}(|\psi_3\rangle|\psi_4\rangle\pm|\psi_4\rangle|\psi_3\rangle),$$
$$\ket{\psi_{\pm}}^{(1,3)}=\frac{1}{\sqrt{2}}(|\psi_1\rangle|\psi_1\rangle\pm|\psi_3\rangle|\psi_3\rangle),\quad
\ket{\phi_{\pm}}^{(1,3)}=\frac{1}{\sqrt{2}}(|\psi_1\rangle|\psi_3\rangle\pm|\psi_3\rangle|\psi_1\rangle),$$
$$\ket{\psi_{\pm}}^{(2,4)}=\frac{1}{\sqrt{2}}(|\psi_2\rangle|\psi_2\rangle\pm|\psi_4\rangle|\psi_4\rangle),\quad
\ket{\phi_{\pm}}^{(2,4)}=\frac{1}{\sqrt{2}}(|\psi_2\rangle|\psi_4\rangle\pm|\psi_4\rangle|\psi_2\rangle),$$
$$\ket{\psi_{\pm}}^{(1,4)}=\frac{1}{\sqrt{2}}(|\psi_1\rangle|\psi_1\rangle\pm|\psi_4\rangle|\psi_4\rangle),\quad
\ket{\phi_{\pm}}^{(1,4)}=\frac{1}{\sqrt{2}}(|\psi_1\rangle|\psi_4\rangle\pm|\psi_4\rangle|\psi_1\rangle),$$
\begin{equation}\label{bell0}
\hspace{1.7cm}\ket{\psi_{\pm}}^{(2,3)}=\frac{1}{\sqrt{2}}(|\psi_2\rangle|\psi_2\rangle\pm|\psi_3\rangle|\psi_3\rangle),\quad
\ket{\phi_{\pm}}^{(2,3)}=\frac{1}{\sqrt{2}}(|\psi_2\rangle|\psi_3\rangle\pm|\psi_3\rangle|\psi_2\rangle)
\end{equation}
and introduce the following $16$ orthonormal entangled states as
follows:
$$\ket{\Phi^1}=\cos\theta\ket{\psi_{+}}^{(1,2)}+\sin\theta\ket{\psi_{+}}^{(3,4)},\quad
\ket{\Phi^2}=-\sin\theta\ket{\psi_{+}}^{(1,2)}+\cos\theta\ket{\psi_{+}}^{(3,4)},$$
$$\ket{\Phi^3}=\cos\theta\ket{\psi_{-}}^{(1,2)}+\sin\theta\ket{\psi_{-}}^{(3,4)},\quad
\ket{\Phi^4}=-\sin\theta\ket{\psi_{-}}^{(1,2)}+\cos\theta\ket{\psi_{-}}^{(3,4)},$$
$$\ket{\Phi^5}=\cos\theta\ket{\phi_{+}}^{(1,2)}+\sin\theta\ket{\phi_{+}}^{(3,4)},\quad
\ket{\Phi^6}=-\sin\theta\ket{\phi_{+}}^{(1,2)}+\cos\theta\ket{\phi_{+}}^{(3,4)},$$
$$\ket{\Phi^7}=\cos\theta\ket{\phi_{-}}^{(1,2)}+\sin\theta\ket{\phi_{-}}^{(3,4)},\quad
\ket{\Phi^8}=-\sin\theta\ket{\phi_{-}}^{(1,2)}+\cos\theta\ket{\phi_{-}}^{(3,4)},$$
$$\hspace{0.2cm}\ket{\Phi^9}=\cos\theta\ket{\phi_{+}}^{(1,3)}+\sin\theta\ket{\phi_{+}}^{(2,4)},\quad
\ket{\Phi^{10}}=-\sin\theta\ket{\phi_{+}}^{(1,3)}+\cos\theta\ket{\phi_{+}}^{(2,4)},$$
$$\hspace{0.2cm}\ket{\Phi^{11}}=\cos\theta\ket{\phi_{-}}^{(1,3)}+\sin\theta\ket{\phi_{-}}^{(2,4)},\quad
\ket{\Phi^{12}}=-\sin\theta\ket{\phi_{-}}^{(1,3)}+\cos\theta\ket{\phi_{-}}^{(2,4)},$$
$$\hspace{0.2cm}\ket{\Phi^{13}}=\cos\theta\ket{\phi_{+}}^{(1,4)}+\sin\theta\ket{\phi_{+}}^{(2,3)},\quad
\ket{\Phi^{14}}=-\sin\theta\ket{\phi_{+}}^{(1,4)}+\cos\theta\ket{\phi_{+}}^{(2,3)},$$
\begin{equation}\label{iso}
\hspace{1.57cm}\ket{\Phi^{15}}=\cos\theta\ket{\phi_{-}}^{(1,4)}+\sin\theta\ket{\phi_{-}}^{(2,3)},\quad\ket{\Phi^{16}}=-\sin\theta\ket{\phi_{-}}^{(1,4)}+\cos\theta\ket{\phi_{-}}^{(2,3)}.
\end{equation}
Although we will not deal with the concurrence of these states,
due to the similarity of these states to the iso-concurrence
states in the two-qubit systems considered in
\cite{akhtar}-\cite{rez}, we refer to these states as
iso-concurrence type states. We note that, for $\theta=\pi/4$ in
(\ref{iso}) we obtain the so-called Bell-type states which are
maximally entangled states. For example we have
$$Tr(W^{(2;4;i_1,...,i_4)}_{opt.}\ket{\Phi^1}\langle\Phi^1|)=1+\sin2\theta[(-1)^{i_1}-(-1)^{i_2}+(-1)^{i_3}-(-1)^{i_4}]-(-1)^{i_1+...+i_4},
$$
\begin{equation}\label{iso0'}
\hspace{-4.2cm}Tr(W'^{(2;4;i_1,i_2;1)}_{opt.}\ket{\Phi^1}\langle\Phi^1|)=1+(-1)^{i_1}+\sin2\theta[(-1)^{i_2}-(-1)^{i_1+i_2}],
\end{equation}
where, the most negative value of (\ref{iso0'}) is obtained for
$\theta=\pi/4$ by taking $i_1=i_3=1,i_2=i_4=0$ in
$W^{(2;4;i_1,...,i_4)}_{opt.}$ and $i_1=i_2=1$ in
$W'^{(2;4;i_1,i_2;1)}_{opt.}$, respectively.

We consider now the spinor ``EPR state'' \cite{pachos} as follows
\begin{equation}\label{EPR}
|\Psi(\vec{P}_1=0,\vec{P}_2=0)\rangle=\sqrt{\frac{m}{2}}(\ket{\psi_4}
\ket{\psi_1}-i\ket{\psi_1} \ket{\psi_4}),
\end{equation}
where, $\vec{P}$ is three-momentum. This state corresponds to a
Lorentz frame where both particles are at rest. As far as the
detection of entanglement is concerned, the Lorentz transformation
do not change the situation, since these transformations take
product states to some another product ones \cite{4} and so
preserves the entanglement. We note that the ``EPR state"
(\ref{EPR}) can be obtained from the state $\ket{\phi_-}^{(1,4)}$ by
applying the rotation $S=e^{i\pi/4 I\otimes\sigma_z}$ on the first
particle. It follows that, if an EW $W$ can detect the state
$\ket{\phi_-}^{(1,4)}$, then $(S\otimes I)W(S\otimes I)^{-1}$ will
be detect the ``EPR state'' $|\Psi(\vec{P}_1=0,\vec{P}_2=0)\rangle$.
Now, one can easily check that
\begin{equation}\label{EPR'0}
Tr(W^{(2;4;i_1,...,i_4)}_{opt.}\ket{\phi_-}^{(1,4)}\langle\phi_-|^{(1,4)})=1-(-1)^{i_1}-(-1)^{i_2}-(-1)^{i_3},
\end{equation}
which shows that $W^{(2;4;0,0,0,i_4)}_{opt.}$, $i_4=0,1$ detect
$\ket{\phi_-}^{(1,4)}$. By taking the similarity transformation
$(S\otimes I)W^{(2;4;0,0,0,i_4)}_{opt.}(S\otimes I)^{-1}$, we
obtain
\begin{equation}\label{EPR'00}
\tilde{W}^{(2;4;0,0,0,i_4)}_{opt.}=(S\otimes
I)W^{(2;4;0,0,0,i_4)}_{opt.}(S\otimes
I)^{-1}=I-\gamma_2\otimes\gamma_1+\gamma_1\otimes\gamma_2+\gamma_3\otimes\gamma_3+(-1)^{i_4}\gamma_4\otimes\gamma_4-(-1)^{i_4}\gamma_5\otimes\gamma_5,
\end{equation}
where, we have used the equalities $S\gamma_1S^{-1}=-\gamma_2$,
$S\gamma_2S^{-1}=\gamma_1$, $S\gamma_3S^{-1}=\gamma_3$,
$S\gamma_4S^{-1}=\gamma_4$, $S\gamma_5S^{-1}=\gamma_5$. Then, one
can easily obtain
$Tr(\tilde{W}^{(2;4;0,0,0,i_4)}_{opt.}\ket{\Psi(\vec{P}_1=0,\vec{P}_2=0)}\langle\Psi(\vec{P}_1=0,\vec{P}_2=0)|)=-2.
$

Now, let $\rho_{_{BSD}}(0)$ be a so called Bell-states diagonal
(BSD) density matrix of a bipartite system in the rest frame which
has the following decomposition
\begin{equation}\label{BD}
\rho_{_{BSD}}(0)=\sum_{i=0}^{15}a_i\ket{\Psi_i(0)}\langle\Psi_i(0)|,\quad
\sum_{i} a_i=1,\quad a_i\geq0
\end{equation}
where, $\ket{\Psi_i(0)}$, denote the Bell-type states obtained by
taking $\theta=\pi/4$ in (\ref{iso}). In the appendix $E$, we show
that any such BSD density matrix can be written as
\begin{equation}\label{BD0}
\rho_{_{BSD}}(0)=\frac{1}{16}I\otimes I+\sum_{\mu=0}^{14}b_\mu
A_{\mu}\otimes A_{\mu},
\end{equation}
where, $A_{\mu}$'s are given by
$$A_{\mu}=\gamma^{\mu};\mu=0,1,2,3\;\ A_4=\gamma^5,\quad A_5=\gamma^0\gamma^1,\;\
A_6=\gamma^0\gamma^2,\;\ A_7=-i\gamma^0\gamma^3,\;\
A_8=i\gamma^1\gamma^2,$$
\begin{equation}\label{amu}
A_{9}=-i\gamma^1\gamma^3, \;\ A_{10}=i\gamma^2\gamma^3,\;\
A_{11}=-i\gamma^0\gamma^5,\;\ A_{12}=\gamma^1\gamma^5,\;\
A_{13}=\gamma^2\gamma^5,\;\ A_{14}=\gamma^3\gamma^5,
\end{equation}
such that $\gamma^{\mu}$ for $\mu=0,1,2,3,5$ defined as
\begin{equation}\label{chiral}
\gamma^0=\sigma_x\otimes I,\;\ \gamma^1=i\sigma_y\otimes
\sigma_x,\;\ \gamma^2=i\sigma_y\otimes \sigma_y,\;\
\gamma^3=i\sigma_y\otimes \sigma_z,\;\ \gamma^5=i\gamma^0\gamma^1
\gamma^2\gamma^3=-\sigma_z\otimes I,
\end{equation}
are the gamma matrices in the chiral representation.

 Clearly, the coefficients $b_\mu$ in (\ref{BD0}) are given by
$b_{\mu}=\frac{1}{16}Tr(\rho_{_{BSD}}(0)A_{\mu}\otimes A_{\mu})$.
The positivity of $\rho_{_{BSD}}(0)$ implies that
$$
\lambda^{i_1,...,i_4}_{_{BD}}=1/16+(-1)^{i_0}b_0+(-1)^{i_1}b_1+(-1)^{i_2}b_2+(-1)^{i_3}b_3-(-1)^{i_0+...+i_3}b_4+(-1)^{i_0+i_1}b_5+(-1)^{i_0+i_2}b_6-$$
$$(-1)^{i_0+i_3}b_7-(-1)^{i_1+i_2}b_8-(-1)^{i_1+i_3}b_{9}-(-1)^{i_2+i_3}b_{10}+(-1)^{i_1+i_2+i_3}b_{11}-(-1)^{i_0+i_2+i_3}b_{12}-$$
\begin{equation}\label{BD00}
(-1)^{i_0+i_1+i_3}b_{13}-(-1)^{i_0+i_1+i_2}b_{14}\geq0,
\end{equation}
Moreover, by imposing the positivity of partial transposition of
$\rho_{_{BSD}}(0)$, we obtain $16$ other inequalities as
(\ref{BD00}) in which the sign of the coefficients
$b_1,b_3,b_{6},b_{9}, b_{11}$ and $b_{13}$ are opposite with those
of (\ref{BD00}). The region defined by intersection of these $32$
halfspaces is a convex polytope which is the region of PPT density
matrices, where its vertices can be obtained by maximizing the left
hand side of one of the inequalities (corresponding to the
halfspaces) subject to the other $31$ inequalities as constraints
(this can be done simply with the simplex method in maple). On the
other hand, the intersection of halfspaces defined by
\begin{equation}\label{BD000}
Tr(\rho_{_{BSD}}(0)W^{(2;4;i_0,...,i_3)}_{opt.})=1+16[(-1)^{i_0}b_0+(-1)^{i_1}b_1+(-1)^{i_2}b_2+(-1)^{i_3}b_3-(-1)^{i_0+...+i_3}b_4]\geq0,
\end{equation}
form a convex polytope, where the intersection of its complement
and the region of PPT density matrices, is the region of
detectable PPT entangled states.

In order to simply determine the region of separable and PPT
entangled states, we consider the special case of BSD density
matrices which are written as
\begin{equation}\label{BSDl}
\rho_{_{BSD}}(0)=\frac{1}{16}I_{16}+b_0\gamma^0\otimes\gamma^0+b_1\gamma^1\otimes\gamma^1+b_2\gamma^2\otimes\gamma^2+b_3\gamma^3\otimes\gamma^3+b_4\gamma^5\otimes\gamma^5.
\end{equation}
Then, the positivity condition (\ref{BD00}) implies that
\begin{equation}\label{ineq00}
\frac{1}{16}+(-1)^{i_0}b_{0}+(-1)^{i_1}b_{1}+(-1)^{i_2}b_{2}+(-1)^{i_3}b_{3}-(-1)^{i_0+...+i_3}b_{4}\geq
0.
\end{equation}
The inequalities (\ref{ineq00}) define a polyhedron in the
$5$-dimensional space with coordinates $(b_0,b_1,b_2,b_3,b_4)$ where
its vertices are as follows
\begin{equation}\label{ineq000}
\rho_{_{BSD}}^{(i_0,...,i_3)}(0)=\frac{1}{48}((-1)^{i_0},(-1)^{i_1},(-1)^{i_2},(-1)^{i_3},-(-1)^{i_0+...+i_3}),\quad
i_1,...,i_4\in\{0,1\}.
\end{equation}

Now, one can easily show that the optimal EWs of the first kind in
the case of $m=2,d=4$ are given by
\begin{equation}\label{optw}
W^{(2;4;i_0,...,i_3)}_{opt.}=I_{16}+
(-1)^{i_0}\gamma^0\otimes\gamma^0+(-1)^{i_1}\gamma^1\otimes\gamma^1+(-1)^{i_2}\gamma^2\otimes\gamma^2+(-1)^{i_3}\gamma^3\otimes\gamma^3+(-1)^{i_0+...+i_3}\gamma^5\otimes\gamma^5.
\end{equation}
By using (\ref{ineq000}) and (\ref{optw}), one can obtain
\begin{equation}\label{tropt}
Tr[W^{(2;4;i_0,...,i_3)}_{opt.}\rho_{_{BSD}}^{(j_0,...,j_3)}(0)]=1+\frac{1}{3}[
(-1)^{i_0+j_0}+(-1)^{i_1+j_1}+(-1)^{i_2+j_2}+(-1)^{i_3+j_3}-(-1)^{i^0+...+i^3+j_0+...+j_3}].
\end{equation}
Then, $\rho_{_{BSD}}^{(j_0,...,j_3)}(0)$ can be detected by
$W^{(2;4;1-j_0,...,1-j_3)}_{opt.}$, where
\begin{equation}\label{tropt1}
Tr[W^{(2;4;1-j_0,...,1-j_3)}_{opt.}\rho_{_{BSD}}^{(j_0,...,j_3)}(0)]=1-\frac{5}{3}=-\frac{2}{3}.
\end{equation}

In order to determine the region of entangled states, we consider
the constraints defined by
\begin{equation}\label{ineqsep}
Tr(\rho_{_{BSD}}(0)W^{(2;4;i_0,...,i_3)}_{opt.}
)=16(\frac{1}{16}+(-1)^{i_0}b_0+(-1)^{i_1}b_1+(-1)^{i_2}b_2+(-1)^{i_3}b_3+(-1)^{i_0+...+i_3}b_4)\geq0
\end{equation}
The inequalities (\ref{ineq00}) and (\ref{ineqsep}) form a
polyhedron with vertices $(\pm1,0,0,0,0)$, $(0,\pm1,0,0,0)$ and
$(0,0,\pm1,0,0)$, $(0,0,0,\pm1,0)$ and $(0,0,0,0,\pm1)$. It should
be noticed that, these density matrices, i.e., $\frac{1}{16}(I\pm
\gamma^i\otimes\gamma^i)$, $i=0,1,2,3,5$ can be written as
superposition of pure product states and hence are separable.
Therefore, the polyhedron defined by $16$ inequalities of
(\ref{ineq00}) is divided to $17$ regions: the central region
which corresponds to the polyhedron defined by (\ref{ineq00}) and
(\ref{ineqsep}), is the region of separable states. The other $16$
regions are in fact the smaller polyhedrons which are associated
with the PPT entangled states. Each of these polyhedrons
corresponds to an offence of one of the $16$ inequalities in
(\ref{ineqsep}).

So far, we considered the BSD density matrices in the rest frame
$S$, where the spinors are at rest. Now, we describe the situation
where, the particles are moving with constant velocity with respect
to each other. We take the spinor representation $D(L(p))$ of a
standard boost $L(p)$ of rapidity $\xi$ as
$$D(L(p))=\exp\{-\frac{\xi}{2}\left(\begin{array}{cc}
                      \vec{\sigma}\cdot \vec{p} & 0 \\
                                        0 & -\vec{\sigma}\cdot\vec{ p} \\
                                      \end{array}\right)\}=$$
\begin{equation}\label{lorentz}
\cosh(\xi/2)\left(\begin{array}{cccc}
              1-p_3\tanh(\xi/2) & -p_-\tanh(\xi/2) & 0 & 0 \\
              -p_+\tanh(\xi/2) & 1+p_3\tanh(\xi/2) & 0 & 0 \\
              0 & 0 & 1+p_3\tanh(\xi/2) & p_-\tanh(\xi/2) \\
              0 & 0 & p_+\tanh(\xi/2) & 1-p_3\tanh(\xi/2) \\
            \end{array}\right).
\end{equation}
In the above equation, $L(p)$ is a coordinate Lorentz
transformation to a frame $S'$ moving with velocity
$v/c=\tanh(-\xi)$ such that from $S'$ the particle at rest in
frame $S$ is observed to have velocity $v/c$. The vector
$\vec{p}=(p_1,p_2,p_3)$ is a unit vector in the direction of
$\mathbf{p}$ with $p_{\pm}=p_1\pm ip_2$.

We are now ready to describe the transformed spinors, by using the
rest frame spinors in Eq.(\ref{helicity}) as follows
\begin{equation}\label{spin}
|\psi_i(p)\rangle=\frac{1}{\sqrt{\cosh
\xi}}D(L(p))|\psi_i(0)\rangle,
\end{equation}
($|\psi_i(0)\rangle\equiv|\psi_i\rangle$ are the helicity basis
defined by Eq.(\ref{helicity})). Then, the rest frame BSD density
matrices given by Eq.(\ref{BD}) are transformed as
$$
\rho_{_{BSD}}(\vec{p})=\frac{1}{\cosh^2(\xi)}[D(L(p))\otimes
D(L(p))]\rho_{_{BSD}}(0)[D^{\dag}(L(p))\otimes
D^{\dag}(L(p))]=$$
\begin{equation}\label{densp}
\sum_{i=0}^{15}a_i\ket{\Psi_i(p)}\langle\Psi_i(p)|, \;\ \sum_{i}
a_i=1,\;\ a_i\geq0.
\end{equation}
In order to avoid more complexities, we consider the BSD density
matrices given by (\ref{BSDl}) with  $\vec{p}=(0,0,1)$. Then the
Lorentz transformation $D(L(p))$ reads
\begin{equation}\label{LT}
D(L(p))=D^{\dag}(L(p))=\cosh(\xi/2)(I\otimes
I-\tanh(\xi/2)\sigma_z\otimes\sigma_z).
\end{equation}
In the following, we discuss the effect of the Lorentz
transformation (\ref{LT}) on the amount of entanglement. To do so,
we will use the Hilbert-Schmidt measure of entanglement
\cite{hilbersh}. In order to define this measure, we recall that the
Hilbert-Schmidt norm is defined as
\begin{equation}\label{HSn}
\|A\|=\sqrt{\langle A,A\rangle},
\end{equation}
where, $\langle A,B\rangle=Tr(A^{\dag}B)$. With help of the norm
(\ref{HSn}), the Hilbert-Schmidt distance between two arbitrary
states $\rho_1,\rho_2$ can be defined as
\begin{equation}\label{HSd}
d_{HS}(\rho_1,\rho_2)=\|\rho_1-\rho_2\|.
\end{equation}
By using the Hilbert-Schmidt distance, the so-called Hilbert-Schmidt
measure of entanglement is defined as
\begin{equation}\label{HS}
D(\rho_{ent.})=\min_{\rho\in S}\|\rho-\rho_{ent.}\|,
\end{equation}
where, $S$ is the set of separable states. In fact, the
Hilbert-Schmidt measure is the minimal distance of an entangled
state $\rho_{ent.}$ to the set of separable states.

For an entangled state $\rho_{ent}$, the minimum of the Hilbert-
Schmidt distance (the Hilbert-Schmidt measure) is attained for some
state $\rho_s$ since the norm is continuous and the set $S$ is
compact. Due to the Bertlmann-Narnhofer-Thirring Theorem \cite{16},
there exist an equivalence between the Hilbert-Schmidt measure and
the concept of optimal entanglement witnesses as follows: The
Hilbert-Schmidt measure of an entangled state equals the maximal
violation of the inequality $Tr(W\rho)\geq 0$,
\begin{equation}\label{HSm}
D(\rho_{_{ent}})=\|\rho_s-\rho_{_{ent}}\|=-\langle\rho_{_{ent}},W_{opt}\rangle=-Tr(\rho_{_{ent}}W_{opt}),
\end{equation}
where,
\begin{equation}\label{HSm1}
W_{opt}=\frac{\rho_s-\rho_{_{ent}}-\langle\rho_{s},\rho_{s}-\rho_{_{ent}}\rangle\mathbf{1}}{\|\rho_s-\rho_{ent.}\|}
\end{equation}
ia an optimal entanglement witness (for more details see Refs.
\cite{hilbersh}, \cite{16}). Therefore, in order to calculate the
Hilbert-Schmidt measure for the PPT BSD entangled states in the rest
frame given by Eq.(\ref{ineq000}), we will use the optimal EWs
(\ref{optw}) and Eq.(\ref{HSm1}) to obtain the state $\rho_s$ in
(\ref{HSm1}). Then, by using the Lorentz transformation (\ref{LT})
we calculate $\rho_{ent}(p)$ and $\rho_s(p)$ which lead us to obtain
the optimal EW for $\rho_{ent}(p)$, by using the Eq.(\ref{HSm1}) and
compute the Hilbert-Schmidt measure for the transformed state
$\rho_{ent}(p)$.

For instance, we consider one of the PPT BSD entangled states given
by (\ref{ineq000}) as
\begin{equation}\label{BSDex}
\rho^{(1,0,0,0)}_{ent.}(0)=\frac{1}{16}\{I\otimes
I-\frac{1}{3}(\gamma^0\otimes\gamma^0-\gamma^1\otimes\gamma^1-\gamma^2\otimes\gamma^2-\gamma^3\otimes\gamma^3-\gamma^5\otimes\gamma^5)\}.
\end{equation}
Then, Eq.(\ref{tropt1}) implies that the optimal EW
\begin{equation}\label{LIW}
W^{(2;4;0,1,1,1)}_{opt.}=I\otimes
I+\gamma^0\otimes\gamma^0-\gamma^1\otimes\gamma^1-\gamma^2\otimes\gamma^2-\gamma^3\otimes\gamma^3-\gamma^5\otimes\gamma^5,
\end{equation}
detects $\rho^{(1,0,0,0)}_{ent.}(0)$. It should be noticed that the
optimal EW (\ref{LIW}) is a Lorentz invariant EW in the sense that
$$
{W^{'(2;4;0,1,1,1)}_{opt.}}=[D^{-1}(L(P))\otimes
D^{-1}(L(P))]W^{(2;4;0,1,1,1)}_{opt.}[D(L(P))\otimes
D(L(P))]=$$
\begin{equation}\label{invar.}
I_{16}+g_{\mu\nu}L(p)_{\alpha}^{\mu}L(p)_{\beta}^{\nu}\gamma^{\alpha}\otimes\gamma^{\beta}-[det(L(P))]^2\gamma^5\otimes\gamma^5=W^{(2;4;0,1,1,1)}_{opt.},
\end{equation}
where, we have used the fact that
$D^{-1}(L(P))\gamma^{\mu}D(L(P))=L(P)_{\nu}^{\mu}\gamma^{\nu}$.

Now, by using (\ref{HSm}) and (\ref{HSm1}), one can write
\begin{equation}\label{Eq}
\rho_s(0)=\rho_{ent}(0)-Tr(\rho_{ent}(0)W^{(2;4;0,1,1,1)}_{opt.})W^{(2;4;0,1,1,1)}_{opt.}+\varepsilon(0)\mathbf{1},
\end{equation}
where,
\begin{equation}\label{Eq1}
\varepsilon(0):=\langle\rho_s(0),\rho_s(0)-\rho_{_{ent}}(0)\rangle=\frac{Tr(\rho_{_{ent}}(0)W^{(2;4;0,1,1,1)}_{opt.})[Tr(W_{opt.}^{(2;4;0,1,1,1)2})-1]}{Tr(W^{(2;4;0,1,1,1)}_{opt.})}.
\end{equation}
In the Eq.(\ref{Eq1}), we have used the optimality of
$W^{(2;4;0,1,1,1)}_{opt.}$ to write
$Tr(\rho_s(0)W^{(2;4;0,1,1,1)}_{opt.})=0$. By substituting
(\ref{BSDex}) and (\ref{LIW}) in (\ref{Eq1}) and using
(\ref{tropt1}), one can obtain $\varepsilon(0)=-\frac{95}{24}$ and
then
\begin{equation}\label{Eq2}
\rho_s(0)=-\frac{31}{48}\{5I\otimes
I-\gamma^0\otimes\gamma^0+\gamma^1\otimes\gamma^1+\gamma^2\otimes\gamma^2+\gamma^3\otimes\gamma^3+\gamma^5\otimes\gamma^5\}.
\end{equation}
The state (\ref{Eq2}) is clearly separable since the states
$I\otimes I\pm \gamma^{\mu}\otimes\gamma^{\mu}$ are product states
for $\mu=0,1,2,3,5$. We normalize the obtained state $\rho_s(0)$ as
\begin{equation}\label{Eq3}
\rho_s(0)=\frac{1}{80}\{5I\otimes
I-\gamma^0\otimes\gamma^0+\gamma^1\otimes\gamma^1+\gamma^2\otimes\gamma^2+\gamma^3\otimes\gamma^3+\gamma^5\otimes\gamma^5\}
\end{equation}
such that $Tr(\rho_s(0))=1$. By this normalization, $\varepsilon(0)$
changes to $\varepsilon(0)=-\frac{1}{120}$ and by using
(\ref{HSm1}), $W^{(2;4;0,1,1,1)}_{opt.}$ is rewritten as
\begin{equation}\label{LIW'}
W^{(2;4;0,1,1,1)}_{opt.}=\frac{1}{4\sqrt{5}}(I\otimes
I+\gamma^0\otimes\gamma^0-\gamma^1\otimes\gamma^1-\gamma^2\otimes\gamma^2-\gamma^3\otimes\gamma^3-\gamma^5\otimes\gamma^5),
\end{equation}
Then, by using (\ref{HSm}) and (\ref{LIW'}), we calculate the
Hilbert-Schmidt measure of $\rho^{(1,0,0,0)}_{_{ent}}(0)$ as
\begin{equation}\label{HSm''}
D(\rho^{(1,0,0,0)}_{_{ent}}(0))=\|\rho_s(0)-\rho^{(1,0,0,0)}_{_{ent}}(0)\|=-Tr(\rho^{(1,0,0,0)}_{_{ent}}(0)W^{(2;4;0,1,1,1)}_{opt.})=\frac{2}{3}.\frac{1}{4\sqrt{5}}=\frac{\sqrt{5}}{30}.
\end{equation}
Now, by using the Lorentz transformation (\ref{LT}), one can
evaluate the transformed states $\rho^{(1,0,0,0)}_{_{ent}}(p)$ and
$\rho_{s}(p)$ as
$$\hspace{-0.65cm}\rho^{(1,0,0,0)}_{_{ent}}(p)=\frac{\cosh^4(\xi/2)}{16\cosh^2(\xi)}\{(1+\tanh^2(\xi/2))^2I_4\otimes I_4+2\tanh(\xi/2)(1+\tanh^2(\xi/2))(I_2\otimes I_2\otimes\gamma^0\gamma^3+\gamma^0\gamma^3\otimes I_2 \otimes I_2)+$$
$$\hspace{-1cm}4\tanh^2(\xi/2)\gamma^0\gamma^3\otimes\gamma^0\gamma^3+\frac{1}{3}[(1-\tanh^2(\xi/2))^2(-\gamma^0\otimes\gamma^0+\gamma^3\otimes\gamma^3)+(1+\tanh^2(\xi/2))^2(\gamma^1\otimes\gamma^1+\gamma^2\otimes\gamma^2+\gamma^5\otimes\gamma^5)-$$
$$2i\tanh(\xi/2)(1+\tanh^2(\xi/2))(\gamma^1\otimes \gamma^2\gamma^5+\gamma^2\gamma^5\otimes \gamma^1-\gamma^2\otimes \gamma^1\gamma^5-\gamma^1\gamma^5\otimes \gamma^2-\gamma^5\otimes\gamma^1\gamma^2-\gamma^1\gamma^2\otimes\gamma^5)-$$
\begin{equation}\label{roent}
4\tanh^2(\xi/2)(\gamma^1\gamma^5\otimes\gamma^1\gamma^5+\gamma^2\gamma^5\otimes\gamma^2\gamma^5+\gamma^1\gamma^2\otimes\gamma^1\gamma^2)]\},
\end{equation}
and
$$\hspace{-0.5cm}\rho_{s}(p)=\frac{\cosh^4(\xi/2)}{80\cosh^2(\xi)}\{5(1+\tanh^2(\xi/2))^2I_4\otimes I_4+10\tanh(\xi/2)(1+\tanh^2(\xi/2))(I_2\otimes I_2\otimes\gamma^0\gamma^3+\gamma^0\gamma^3\otimes I \otimes I)+$$
$$\hspace{-0.65cm}20\tanh^2(\xi/2)\gamma^0\gamma^3\otimes\gamma^0\gamma^3+(1-\tanh^2(\xi/2))^2(-\gamma^0\otimes\gamma^0+\gamma^3\otimes\gamma^3)+(1+\tanh^2(\xi/2))^2(\gamma^1\otimes\gamma^1+\gamma^2\otimes\gamma^2+\gamma^5\otimes\gamma^5)-$$
$$2i\tanh(\xi/2)(1+\tanh^2(\xi/2))(\gamma^1\otimes \gamma^2\gamma^5+\gamma^2\gamma^5\otimes \gamma^1-\gamma^2\otimes \gamma^1\gamma^5-\gamma^1\gamma^5\otimes \gamma^2-\gamma^5\otimes\gamma^1\gamma^2-\gamma^1\gamma^2\otimes\gamma^5)-$$
\begin{equation}\label{ros}
4\tanh^2(\xi/2)(\gamma^1\gamma^5\otimes\gamma^1\gamma^5+\gamma^2\gamma^5\otimes\gamma^2\gamma^5+\gamma^1\gamma^2\otimes\gamma^1\gamma^2)\},
\end{equation}
respectively. Then, we obtain
$$\varepsilon(p)=\langle\rho_{s}(p),\rho_{s}(p)-\rho^{(1,0,0,0)}_{ent}(p)\rangle=Tr[\rho_{s}(p)(\rho_{s}(p)-\rho^{(1,0,0,0)}_{ent}(p))]=$$
\begin{equation}\label{epsilonp}
-\frac{\cosh^8(\xi/2)}{600\cosh^4(\xi)}\{5(1+\tanh^8(\xi/2))+28(\tanh^2(\xi/2)+\tanh^6(\xi/2))+126\tanh^4(\xi/2)\}.
\end{equation}
Then, by using (\ref{HSm1}), we obtain the optimal EW associated
with $\rho^{(1,0,0,0)}_{_{ent}}(p)$ as
$$W_{opt}(p)=\frac{\cosh^4(\xi/2)}{120\cosh^2(\xi)\|\rho_{s}(p)-\rho^{(1,0,0,0)}_{_{ent}}(p)\|}\{\frac{180\cosh^2(\xi)}{\cosh^4(\xi/2)}\|\rho_{s}(p)-\rho^{(1,0,0,0)}_{_{ent}}(p)\|^2I\otimes
I+$$
$$(1-\tanh^2(\xi/2))^2(\gamma^0\otimes\gamma^0-\gamma^3\otimes\gamma^3)-(1+\tanh^2(\xi/2))^2(\gamma^1\otimes\gamma^1+\gamma^2\otimes\gamma^2+\gamma^5\otimes\gamma^5)-$$
$$2i\tanh(\xi/2)(1+\tanh^2(\xi/2))^2(-\gamma^1\otimes \gamma^2\gamma^5-\gamma^2\gamma^5\otimes \gamma^1+\gamma^2\otimes \gamma^1\gamma^5+\gamma^1\gamma^5\otimes \gamma^2+\gamma^5\otimes\gamma^1\gamma^2+\gamma^1\gamma^2\otimes\gamma^5)+$$
\begin{equation}\label{Wp}
4\tanh^2(\xi/2)(\gamma^1\gamma^5\otimes\gamma^1\gamma^5+\gamma^2\gamma^5\otimes\gamma^2\gamma^5+\gamma^1\gamma^2\otimes\gamma^1\gamma^2)\}.
\end{equation}
In the appendix $E$, we show that $W_{opt}(p)$ in (\ref{Wp}) is an
EW. Now, we evaluate the Hilbert-Schmidt measure of
$\rho^{(1,0,0,0)}_{_{ent}}(p)$ as follows
$$D(\rho^{(1,0,0,0)}_{_{ent}}(p))=\|\rho_{s}(p)-\rho^{(1,0,0,0)}_{_{ent}}(p)\|=-Tr(W_{opt}(p)\rho^{(1,0,0,0)}_{_{ent}}(p))=$$
\begin{equation}\label{final}
\frac{\cosh^4(\xi/2)}{30\cosh^2(\xi)}\{\sqrt{5(1+\tanh^8(\xi/2))+28(\tanh^2(\xi/2)+\tanh^6(\xi/2))+126\tanh^4(\xi/2)}\;\
\}.
\end{equation}
The above result indicates that the Hilbert-Schmidt measure of
$\rho^{(1,0,0,0)}_{_{ent}}(p)$ is larger than $\frac{\sqrt{5}}{30}$
which is the same as the Hilbert-Schmidt measure of
$\rho^{(1,0,0,0)}_{ent}(0)$, i.e.,
$D(\rho^{(1,0,0,0)}_{_{ent}}(p)\geq
D(\rho^{(1,0,0,0)}_{_{ent}}(0))$. Therefore, as far as the spin
quantum correlations is concerned, the amount of entanglement is not
a relativistic scalar and has no invariant meaning. This result can
be compared with the result of Peres, et. al. in Ref. \cite{peres'},
where it has been shown that the entropy of the reduced density
matrix describing just the spin of a particle (without the momentum)
is not Lorentz invariant. In fact, the result (\ref{final})
indicates that the minimum value of the amount of spin entanglement
of a spin entangled BSD density matrix is archived in the rest
frame.

Now, we return to the rest frame and discus the BSD density matrices
which can be detected via the optimal EWs of the second kind (with
$j=1$). The case of the moving frame can be considered similar to
the above discussions for the EWs of the first kind. In the rest
frame, the optimal EWs of the second kind and the BSD density
matrices are defined as
$$\hspace{-1.5cm}W'^{(2;4;i_1,i_2;1)}_{opt.}=I_{16}+
(-1)^{i_1}A'_1\otimes A'_1+(-1)^{i_2}A'_3\otimes
A'_3+(-1)^{i_1+i_2}A'_5\otimes A'_5,$$
$$
\hspace{1cm}\rho'_{_{BSD}}=\frac{1}{16}I_{16}+ b'_{1}A'_1\otimes
A'_1+b'_{2}A'_2\otimes A'_2+b'_{3}A'_3\otimes
A'_3+b'_{4}A'_4\otimes A'_4+b'_{5}A'_5\otimes
A'_5+b'_{6}A'_6\otimes A'_6
$$
respectively, where $A'_1=i\gamma^1\gamma^2$, $A'_2=\gamma^5$,
$A'_3=\gamma^1\gamma^5$, $A'_4=-i\gamma^2$, $A'_5=\gamma^2\gamma^5$
and $A'_6=-i\gamma^1$. The positivity of $\rho'_{_{BSD}}$ implies
that
\begin{equation}\label{ineq1'}
\frac{1}{16}+(-1)^{i_1}b'_{1}+(-1)^{i_2}b'_{2}+(-1)^{i_3}b'_{3}+(-1)^{i_1+i_2+i_3}b'_4-(-1)^{i_1+i_3}b'_{5}-(-1)^{i_2+i_3}b'_{6}\geq0
\end{equation}
These inequalities define a tetrahedron in the $6$-dimensional space
with coordinates $(b'_1,...,b'_6)$ where its vertices are
$\frac{1}{16}((-1)^{i_1},0,(-1)^{i_2},0,-(-1)^{i_1+i_2},0)$ with
$i_1,i_2\in \{0,1\}$. In order to determine the region of entangled
states, we consider the constraints defined
by
\begin{equation}\label{ineq1'''}
Tr(\rho'_{_{BSD}}W'^{(2;4;i_1,i_2;1)}_{opt.}
)=16(\frac{1}{16}+(-1)^{i_1}b'_1+(-1)^{i_2}b'_3+(-1)^{i_1+i_2}b'_5)\geq0.
\end{equation}
The inequalities (\ref{ineq1'}) and (\ref{ineq1'''}) form a
polytope with vertices $\frac{1}{16}(\pm1,0,0,0,0,0)$,\\
$\frac{1}{16}((-1)^{i_1},0,0,(-1)^{i_2},0,-(-1)^{i_1+i_2})$,
$\frac{1}{16}(0,(-1)^{i_1},(-1)^{i_2},0,0,-(-1)^{i_1+i_2})$ and\\
$\frac{1}{16}(0,(-1)^{i_1},0,(-1)^{i_2},-(-1)^{i_1+i_2},0)$.
Therefore, the polytope defined by inequalities of (\ref{ineq1'})
is divided to five regions: the central region which corresponds
to the octahedron, is the region of separable states. The other
four regions which are all equivalent are in fact the smaller
tetrahedrons which are associated with the entangled states. Each
of these tetrahedrons corresponds to an offence of one of the
inequalities in (\ref{ineq1'''}).
\subsection{Non-relativistic entangled states which can be detected by $W^{(m;d;i_1,...,i_d)}_{opt.}$}
Now, we consider the multispinor systems with density matrices of
the form
\begin{equation}\label{Gdensitys}
\rho_{_{BSD}}^{(m;d)}:=b_0I_{2^{m
d/2}}+\sum_{\mu=1}^{2^d-1}b_\mu\underbrace{A_\mu\otimes...\otimes
A_\mu}_m
\end{equation}
as a generalization of BSD density matrices to the cases of
multipartite and higher dimensional systems, where $A_{\mu}$'s are
hermitian operators obtained by all possible multiplications of
$\gamma^{(d)}_i$, $i=1,...,2d$ as before. The determination of the
region of PPT entangled states detectable by
$W^{(m;d;i_1,...,i_d)}_{opt.}$ is similar to the case of the
bipartite four-dimensional space-time.

We consider the following particular density matrices
\begin{equation}\label{densitys}
\rho_{_{BSD}}^{(m;d)}:=b_0I_{2^{m
d/2}}+\sum_{i=1}^{d+1}b_i\underbrace{\gamma^{(d)}_i\otimes...\otimes
\gamma^{(d)}_i}_m
\end{equation}
Due to tracelessness of $\gamma^{(d)}_i$, the condition
$Tr(\rho_{_{BSD}}^{(m;d)})=1$ gives $b_0=\frac{1}{2^{md/2}}$ and
the positivity of $\rho_{_{BSD}}^{(m;d)}$ imposes the constraints
\begin{equation}\label{pos}
\frac{1}{2^{md/2}}+\sum_{k=1}^{d}(-1)^{i_k}b_k+(-i)^{md/2}(-1)^{i_1+...+i_{d}}b_{d+1}\geq0\quad
,\quad \forall \ (i_{1},i_{2},...,i_{d})\in \{0,1\}^{d}
\end{equation}
to its eigenvalues. The intersection of these $2^{d}$ halfspaces
form a simplex polygon in a $d+1$ dimensional space with
coordinate variables $b_i$ (excepted $b_0$). Furthermore if we
want $\rho_{_{BSD}}^{(m;d)}$ becomes a PPT entangled state in the
sense that its partial transpose is positive definite with respect
to each subsystem, then we will obtain additional constraints
which must be satisfied. For instance, the positivity of the
partial transpose with respect to any particle, i.e.,
${\rho_{_{BSD}}^{(m;d)}}^{T_{i}} \geq 0$, $i=1,2,...,m$ imposes
the following constraints
$$
\frac{1}{2^{md/2}}+ \sum_{k=0}^{d/2-1}(-1)^{i_{2k+1}}
b_{2k+1}-\sum_{k=1}^{d/2}(-1)^{
i_{2k}}b_{2k}+(-i)^{md/2}(-1)^{i_1+...+i_{d}}b_{d+1}\geq0\quad ,
\forall \ (i_{1},i_{2},...,i_{d})\in \{0,1\}^{d}
$$
where, we have used the fact that all $\gamma$ matrices with odd
index are symmetric and all matrices with even index are
antisymmetric (see appendix $A$). In order to determine the region
of PPT entangled states, we consider the constraints obtained by
\begin{equation}\label{pos'}
Tr(\rho_{_{BSD}}^{(m;d)}W_{opt.}^{(m;d;i_1,...,i_d)} )=1+
2^{md/2}(\sum_{k=1}^{d}(-1)^{i_k}
b_{k}-(-i)^{md/2}(-1)^{i_1+...+i_{d}}b_{d+1})\geq0.
\end{equation}
The inequalities (\ref{pos}) and (\ref{pos'}) form a polyhedron
with vertices $(\pm1,0,...,0)$, $(0,\pm1,0,...,0)$,...,
$(0,...,0,\pm1)$ which are the same as the vertex points of the
feasible region. Therefore, the polyhedron defined by inequalities
of (\ref{pos}) is divided to $2^{d}+1$ regions: the central region
which is defined by (\ref{pos}) and (\ref{pos'}), corresponds to
the region of separable states. The other $2^{d}$ regions  are in
fact the smaller polyhedrons which are associated with the PPT
entangled states. Each of these polyhedrons corresponds to an
offence of one of the inequalities in (\ref{pos'}).

We note that, in the the case of $d=m=2$ with
$$W^{(2;2;i_1,i_2)}_{opt.}=I_{4}+
(-1)^{i_1}\sigma_x\otimes\sigma_x+(-1)^{i_2}\sigma_y\otimes\sigma_y+(-1)^{i_1+i_2}\sigma_z\otimes\sigma_z,$$
$$
\hspace{-3.5cm}\rho_{_{BSD}}^{(2;2)}=\frac{1}{4}I_{4}+
b_{1}\sigma_x\otimes\sigma_x+b_{2}\sigma_y\otimes\sigma_y+b_{3}\sigma_z\otimes\sigma_z,
$$
the Eq.(\ref{pos}) implies that
\begin{equation}\label{ineq0}
\frac{1}{4}+(-1)^{i_1}b_{1}+(-1)^{i_2}b_{2}-(-1)^{i_1+i_2}b_{3}\geq
0.
\end{equation}
These inequalities define a tetrahedron in the $3$-dimensional
space with coordinates $(b_1,b_2,b_3)$ where its vertices are
$(-\frac{1}{4},-\frac{1}{4},-\frac{1}{4}),(\frac{1}{4},\frac{1}{4},-\frac{1}{4}),(\frac{1}{4},-\frac{1}{4},\frac{1}{4})$
and $(-\frac{1}{4},\frac{1}{4},\frac{1}{4})$. In order to
determine the region of entangled states, we consider the
constraints obtained by
\begin{equation}\label{ineq0'}
Tr(\rho_{_{BSD}}^{(2;2)}W^{(2;2;i_1,i_2)}_{opt.}
)=4(\frac{1}{4}+(-1)^{i_1}b_1+(-1)^{i_2}b_2-(-1)^{1+i_1+i_2}b_3)\geq0.
\end{equation}
The inequalities (\ref{ineq0}) and (\ref{ineq0'}) form an
octahedron with vertices $(\pm1,0,0)$, $(0,\pm1,0)$ and
$(0,0,\pm1)$. Therefore, the tetrahedron defined by inequalities
of (\ref{ineq0}) is divided to five regions: the central region
which corresponds to the octahedron, is the region of separable
states. The other four regions which are all equivalent are in
fact the smaller tetrahedrons which are associated with the PPT
entangled states. Each of these tetrahedrons corresponds to an
offence of one of the inequalities in (\ref{ineq0'}). It is
interesting to note that, in this case the nonnegativity of
$Tr(\rho_{_{BSD}}^{(2;2)}W^{(2;2;i_1,i_2)}_{opt.})<0$ is
equivalent to $C(\rho_{_{BSD}}^{(2;2)})>0 $, where $C$ is the
concurrence introduced by Wootters \cite{woot}.
\subsection{Non-relativistic entangled states which can be detected by $W'^{(m;d;i_1,i_2;j)}_{opt.}$}
Now we assert that $W'^{(m;d;i_1,i_2;j)}_{opt.}$ given by
(\ref{W'n}) can also detect some entangled multispinor mixed
density matrices. To this aim we consider only the following
density matrices
\begin{equation}\label{density}
{\rho'}_{_{BSD}}^{(m;d)}:=b'_0I_{2^{md/2}}+\sum_{i=1}^{3
d/2}b'_i\underbrace{A'_i\otimes A'_i\otimes...\otimes A'_i}_{m},
\end{equation}
with $b'_0=\frac{1}{2^{md/2}}$, where the general density matrices
as in (\ref{Gdensitys}) can be considered similarly. The
positivity of density matrix ${\rho'}_{_{BSD}}^{(m;d)}$ imposes
\begin{equation}\label{posdensity}
b'_0 +\sum_{k=1}^{d/2+1}(-1)^{i_k}
b'_k+\sum_{k=2}^{d/2}(-1)^{i_1+i_{d/2+1}+i_{k}}b'_{d/2+k}+\sum_{k=1}^{d/2}(-1)^{m/2+i_{d/2+1}+i_{k}}b'_{d+k}\geq0\quad
\end{equation}
with $i_{1},...,i_{d/2+1}\in \{0,1\}$, to its eigenvalues. The
intersection of these halfspaces form a simplex polygon in a
$3d/2$ dimensional space with coordinate variables $b'_i$,
$i=1,...,3d/2$ (excepted $b_0$). The condition for detectability
of ${\rho'}_{_{BSD}}^{(m;d)}$ by $W'^{(m;d;i_1,i_2;j)}_{opt.}$ can
be written as
\begin{equation}\label{eqqq}
Tr({\rho'}_{_{BSD}}^{(m;d)}W'^{(m;d;i_1,i_2;j)}_{opt.})=1+2^{md/2}((-1)^{i_1}b'_j+(-1)^{i_{2}}b'_{d/2+j}-(-1)^{m/2+i_1+i_2}b'_{d+j})<0.
\end{equation}
As we will show in the following section, the EWs
$W'^{(m;d;i_1,i_2;j)}_{opt.}$ are decomposable EWs and so can not
detect PPT entangled states.
\section{Decomposability or non-decomposability of BSD multispinor EWs}
Another interesting feature of EWs is their decomposability or
non-decomposability. Clearly d-EW can not detect PPT entangled
states (these states are also called bound entangled states because
they have the peculiar property that no entanglement can be
distilled from them by local operations \cite{horo}) whereas there
are some bound entangled states which can be detected by a nd-EW. In
the previous section, it was shown that there exist some bound
entangled states which can be detected by the optimal EWs of the
first kind, whereas the EWs of the second kind can not detect bound
entangled states. In fact, the detectability or non-detectability of
bound entangled states is due to non-decomposability or
decomposability of the corresponding EWs where in the following, we
discuss this particular property of the optimal EWs of the first and
second kinds.
\subsection{The region of non-decomposable EWs of the first kind}
First consider the first kind of BSD multispinor EWs $W^{(m;d)}$.
The inequalities of Eq.(\ref{a2}) show that in the space of
parameters $a_i$, all of these EWs lie inside the hypercube (by
fixing $a_{_{0}}$). The operator $W^{(m;d)}$ is positive in the
region defined by the following inequalities
\begin{equation}\label{ineq1}
a_{0}+\sum_{k=1}^{d}(-1)^{i_{k}}a_{k}+(-i)^{md/2}(-1)^{i_1+...+i_{d}}
a_{d+1}\geq 0 \quad (i_{1},...,i_{d})\in\{0,1\}^{d}.
\end{equation}
Now, consider the $2^{d}$ coordinates
$(a_{1},...,a_{d},a_{d+1})\in\{((-1)^{i_1},...,(-1)^{i_{d}},-(-i)^{md/2}(-1)^{i_1+...+i_{d}}):
(i_{1},...,i_{d})\in\{0,1\}^{d}\}$. These coordinates correspond to
the optimal EWs given by Eq.(\ref{region}). The partial
transpositions of the optimal EWs can be written as
$$W^{(m;d;i_1,...,i_{d})T_i}_{opt.}=I-\sum_{j=1}^{d}(-1)^{i_{2j}}\gamma^{(d)}_{2j}\otimes...\otimes
\gamma^{(d)}_{2j}+\sum_{j=1}^{d}(-1)^{i_{2j-1}}\gamma^{(d)}_{2j-1}\otimes...\otimes
\gamma^{(d)}_{2j-1}-$$
\begin{equation}\label{region''}
(-i)^{md/2}(-1)^{i_1+...+i_{d}}\gamma^{(d)}_{d+1}\otimes...\otimes
\gamma^{(d)}_{d+1},
\end{equation}
for $i=1,2,...,m$. Note that, $W^{(m;d;i_1,...,i_{d})T_i}_{opt.}$
are not positive for $d\neq 2$. In fact, the eigenvalues of
$W^{(m;d;i_1,...,i_{d})T_i}_{opt.}$ are given by
\begin{equation}\label{regioneig}
\lambda^{(m;d;i_1,...,i_{d})}_{k_1,...,k_{d}}=1-\sum_{j=1}^{d/2}\sum_{l=1}^{d/2}(-1)^{i_{2j}}(-1)^{k_{2l}}+\sum_{j=1}^{d/2}\sum_{l=1}^{d/2}(-1)^{i_{2j-1}}(-1)^{k_{2l-1}}-(-i)^{md}(-1)^{i_1+...+i_{d}+k_1+...+k_{d}}
\end{equation}
which are not positive with respect to none of the particles. For
example, consider the case $i_1=i_2=...=i_{d}=0$. Then the
eigenvalues will be
\begin{equation}\label{regioneig'}
\lambda^{(m;d;0,...,0)}_{k_1,...,k_{d}}=1-\sum_{l=1}^{d/2}(-1)^{k_{2l}}+\sum_{l=1}^{d/2}(-1)^{k_{2l-1}}-(-1)^{k_1+...+k_{d}},
\end{equation}
where, the most negative eigenvalue is obtained by taking $k_{2l}=0,
k_{2l-1}=1$, $l=1,2,...,d/2$, i.e., we have
\begin{equation}\label{regioneig''}
\lambda^{(m;d;0,...,0)}_{k_{2l}=0, k_{2l-1}=1}=1-d-(-1)^{d/2}\leq
2-d<0,\quad \mathrm{for}\quad d>2.
\end{equation}
For $d=2$, the EW is the same as the EW of the second kind and we
have
\begin{equation}\label{regioneig'''}
W^{(m;2;i_1,i_{2})T_i}_{opt.}=I+(-1)^{i_1}\gamma_1\otimes...\otimes\gamma_1-(-1)^{i_2}\gamma_2\otimes...\otimes\gamma_2-(-1)^{m/2+i_1+i_2}\gamma_3\otimes...\otimes\gamma_3,
\end{equation}
which is positive for all values of $m$ and $i_1,i_2\in\{0,1\}$.

We discuss the non-decomposability of EWs only in the case of
$d=4$, $m=2$, the multipartite case can be discussed similarly.
Now we consider the vertices of the density matrices' region given
by (\ref{ineq000}) (recall that, all of these density matrices are
PPT). In order to determine the region of non-decomposable EWs, we
take the constraints obtained by
\begin{equation}\label{ineq00'}
Tr(\rho^{(2;4)}_{i_1,...,i_4}W^{(2;4)}
)=16(1-\frac{1}{3}[(-1)^{i_1}a_1+(-1)^{i_2}a_2+(-1)^{i_3}a_3+(-1)^{i_4}a_4-(-1)^{i_1+...+i_4}a_5])<0
\end{equation}
which is equivalent to
\begin{equation}\label{ineq0sp}
(-1)^{i_1}a_1+(-1)^{i_2}a_2+(-1)^{i_3}a_3+(-1)^{i_4}a_4-(-1)^{i_1+...+i_4}a_5>3.
\end{equation}
It can be seen that the minimum value of
$1-\frac{1}{3}[(-1)^{i_1}a_1+(-1)^{i_2}a_2+(-1)^{i_3}a_3+(-1)^{i_4}a_4-(-1)^{i_1+...+i_4}a_5]$
is obtained by choosing the parameters $(a_1,...,a_5)$ as
$((-1)^{i_1},(-1)^{i_2},(-1)^{i_3},(-1)^{i_4},$\\$-(-1)^{i_1+...+i_4})$,
which are the same as the optimal EWs, i.e., all of the optimal
EWs $W^{(2;4;i_1,...,i_4)}_{opt.}$ are non-decomposable. Then, we
will have
\begin{equation}\label{ineq00'''}
\mathrm{min}_{_{a_1,...,a_5}}(1-\frac{1}{3}[(-1)^{i_1}a_1+(-1)^{i_2}a_2+(-1)^{i_3}a_3+(-1)^{i_4}a_4-(-1)^{i_1+...+i_4}a_5])=1-\frac{5}{3}=-\frac{2}{3}.
\end{equation}
In fact, the EWs $W^{(2;4)}$ satisfying the inequalities
(\ref{ineq0sp}) are non-decomposable EWs.
\subsection{Decomposability of EWs of the second kind}
Now, consider the second kind of BSD multisinor EWs. In the space
of parameters $ a'_i$ (again by fixing $a'_{0}$), all of these EWs
lie inside the region defined by Eq. (\ref{vp2'}). The region
\begin{equation}\label{ineq1x}
a'_0 +\sum_{k=1}^{d/2+1}(-1)^{i_k}
a'_k+\sum_{k=2}^{d/2}(-1)^{i_1+i_{d/2+1}+i_{k}}a'_{d/2+k}+\sum_{k=1}^{d/2}(-1)^{m/2+i_{d/2+1}+i_{k}}a'_{d+k}\geq
0 \quad (i_{1},...,i_{d})\in\{0,1\}^{d}
\end{equation}
is the place where the EW is positive. Now consider the
coordinates
$(a'_1,...,a'_{d},a'_{d+1})\in((-1)^{i_1},0,...,0,(-1)^{i_2},0,...,0,(-1)^{i_1+i_2})$.
These parameters correspond to the optimal EWs given by
(\ref{W'n}) for $j=1$ (the discussions for $j=2,...,d/2$ are
similar). The partial transpose of these optimal EWs with respect
to each particle is given by
\begin{equation}\label{ineq2'pt}
W'^{(m;d;i_1,i_2;1)T_i}_{opt.}=I+(-1)^{i_1}A'_1\otimes...\otimes
A'_1+(-1)^{i_2}A'_{d/2+1}\otimes...\otimes
A'_{d/2+1}+(-1)^{m/2+i_1+i_2}A'_{d+1}\otimes...\otimes A'_{d+1},
\end{equation}
Then, the eigenvalues of $W'^{(m;d;i_1,i_2;1)T_i}_{opt.}$ are
\begin{equation}\label{ineq2'pteig}
\lambda^{'(m;d;i_1,i_2)}_{k_1,k_2}=1+(-1)^{i_1+k_1}+(-1)^{i_2+k_2}+(-1)^{i_1+i_2+k_1+k_2},
\end{equation}
where, we have used the fact that $A'_{d+1}=iA'_{d/2+1}A'_1$.
Then, one can easily check that
$\lambda^{'(m;d;i_1,i_2)}_{k_1,k_2}$ are positive for all values
of $i_1,i_2,k_1,k_2\in\{0,1\}$. Therefore, the EWs defined by
(\ref{W'n}) have positive partial transpose with respect to each
particle and so are optimal decomposable EWs. A cone which may be
formed by connecting every four points of Eq.(\ref{W'n}) to its
opposite positive hyperplane in Eq.(\ref{ineq1x}) is d-EWs. Note
that the remaining operators in Eq.(\ref{EW'}) coming from some
points in the space of parameters are either d-EW or positive. In
fact, from the convexity of the EWs' region, every EW is written
as a convex combination of the decomposable optimal EWs (the
vertices of the EWs' region) and so is also decomposable.
Therefore we conclude that all of the multispinor EWs of the
second kind are decomposable and can not detect PPT entangled
states.
\section{Multispinor EWs which can be manipulated approximately by LP}
So far, we have considered the BSD multispinor EWs which can be
constructed by the exact LP method, while in this section, we
consider the EWs that can be manipulated by approximate LP which
come from by adding other members of Dirac $\gamma$ matrices
algebra to exactly soluble multispinor EWs. In all of the
multispinor EWs discussed in section $3$, the boundary hyperplanes
arise from the vertex points which themselves come from pure
product states and the resulting inequalities did not offend
against the convex hull of the vertices at all. But  by adding
some terms to exactly soluble EWs, it may be happen that the
feasible region be convex with curvature on some boundaries and
the problem can not be solved by the exact LP method. In these
cases the linear constraints no longer arise from convex hull of
the vertices coming from pure product states. Hence we transform
such problem to the approximate LP one. Our approach is to draw
the hyperplanes tangent to feasible region and parallel to
hyperplanes coming from vertices and in this way we enclose the
feasible regions by such hyperplanes. It is clear that in this
extension, the vertices no longer arise from pure product states.
\subsection{The first kind}
In the case of the first kind of BSD multispinor EWs, we add one
of the multiplication of the matrices $\gamma_i$, $i=1,2,...,d+1$,
say, $-i\gamma_1\gamma_2$ to (\ref{Wn}) as
\begin{equation}\label{apWn}
W_{ap.}^{(m;d)}=a_{0}I_{2^{md/2}}+\sum_{k=1}^{d+1}
a_{k}\gamma^{(d)}_k\otimes...\otimes\gamma^{(d)}_k+(-i)^ma_{d+2}\gamma^{(d)}_1\gamma^{(d)}_2\otimes...\otimes\gamma^{(d)}_1\gamma^{(d)}_2.
\end{equation}
(the subscript ap. refers to the approximate) and try to solve it
by LP method. The eigenvalues of $W_{ap.}^{(m;d)}$ are
$$
\lambda^{(m;d)}_{ap.;i_1,...,i_{d}}=a_{0}+\sum_{k=1}^{d}
(-1)^{i_{k}} a_{_{k}}+i^{-md/2}(-1)^{i_{1}+...
i_{d}}a_{d+1}+(-i)^m(-1)^{i_{1}+ i_{2}}a_{d+2}\quad ,\quad \forall
\ (i_{1},i_{2},...,i_{d})\in \{0,1\}^{d}
$$
The coordinates of the apexes which arise from pure product states
are listed in the following table
\begin{equation}\label{tab1}
\begin{tabular}{c|c}\hline\hline
  Product state & $(P_{1},P_{2},...,P_{d},P_{d+1},P_{d+2})$ \\ \hline
  $|\psi^{(1)}_{\pm}\rangle$ & $(\pm1,0,0,...,0) $ \\
  $|\psi^{(2)}_{\pm}\rangle$ & $(0,\pm1,0,...,0)$ \\
  $\vdots$ & $\vdots$ \\
  $|\psi^{(d+1)}_{\pm}\rangle$ & $(0,...,0,\pm1,0)$ \\
  $|\psi^{(d+2)}_{\pm}\rangle$ & $(0,0,...,0,\pm1)$,\\
         \hline\hline
\end{tabular}
\end{equation}
where, $|\psi^{(i)}_{\pm}\rangle$ for $i=1,...,d+1$ are defined as
in section $3.2.1$ and $|\psi^{(d+2)}_{\pm}\rangle$ are
eigenvectors of
$(-i)^m\gamma^{(d)}_1\gamma^{(d)}_2\otimes...\otimes\gamma^{(d)}_1\gamma^{(d)}_2$
with eigenvalues $\pm1$. Then, the feasible region is the
intersection of the following halfspaces
\begin{equation}\label{ineapprev}
\sqrt{2}+\sum_{k=1}^{d+2}(-1)^{i_{k}}P_{k}\geq 0,
\end{equation}
where $i_{1},...,i_{d+2}\in \{0,1\}$ (the proof of (\ref{ineapprev})
is given in appendix $D$). The inequalities (\ref{ineapprev}) imply
that the problem does not lie in the realm of exactly soluble LP
problems and we have to use approximate LP. To this end we shift
aforementioned hyperplanes parallel to themselves such that they
reach to maximum value $\sqrt{2}$. On the other hand the maximum
shifting is where the hyperplanes become tangent to convex region
coming from pure product states and in this manner we will be able
to encircle the feasible region by the hyperspaces defined by
(\ref{ineapprev}).

Regarding the above considerations, the problem is reduced to
$$
\hspace{-3.6cm}\mathrm{minimize} \quad\;\;\;\;\;\
a_0+\sum_{i=1}^{d+2}a_iP_i\vspace{-1mm}
$$
\begin{equation}\label{region1ap}
\hspace{-2cm} \mathrm{subject\ to}
\quad\;\;\left\{\begin{array}{c}
 \sqrt{2}+\sum_{k=1}^{d+2}(-1)^{i_{k}}P_{k}\geq0
\\
 \hspace{-2cm}\forall\ |P_k|\leq 1,\\
\end{array}\right.
\end{equation}
for all $i_{1},...,i_{d+2}\in \{0,1\}$, where it can be solved by
simplex method, since the intersections of the hyperspaces in
(\ref{region1ap}) form a convex polytope.

By substitution of extreme points of the feasible region (we note
that these points do not arise from pure product states), we get
the approximate region of SSNNEV as intersection of the following
halfspaces
\begin{equation}\label{apa}
|a_i|\leq \frac{1}{\sqrt{2}}a_0,\quad i=1,...,d+1,d+2.
\end{equation}
In fact, the approximated region of EWs is the complement of the
region defined by $\lambda^{(m;d)}_{ap.;i_1,...,i_{d}}\geq0$ in
the hypercube defined by (\ref{apa}).
\subsubsection{The region of non-decomposable (approximate) EWs of
the first kind} The inequalities of Eq.(\ref{apa}) show that in
the space of parameters $a_i$, all of the EWs $W_{ap.}^{(m;d)}$
lie inside a hypercube (by fixing $a_{_{0}}$). Also, these EWs are
positive in the region defined by the following inequalities
\begin{equation}\label{ineq1}
a_{0}+\sum_{k=1}^{d}(-1)^{i_{k}}a_{k}+(-i)^{md/2}(-1)^{i_1+...+i_{d}}
a_{d+1}+(-i)^m(-1)^{i_1+i_2}a_{d+2}\geq 0 \quad
(i_{1},...,i_{d})\in\{0,1\}^{d}.
\end{equation}
 Now consider the coordinates
$(a_{1},...,a_{d},a_{d+1},a_{d+2})\in\{((-1)^{i_1},(-1)^{i_{2}},0,0,...,0,(-1)^{i_1+i_2}):
i_{1},i_{2}\in\{0,1\}\}$. Substituting these $2^{d}$ points in
$W_{ap.}^{(m;d)}$ gives the optimal EWs in the approximated region
as follows
\begin{equation}\label{region'ap}
W^{(m;d;i_1,...,i_{d})}_{ap.,opt.}=I+(-1)^{i_1}\gamma^{(d)}_1\otimes...\otimes
\gamma^{(d)}_1+(-1)^{i_2}\gamma^{(d)}_{2}\otimes...\otimes
\gamma^{(d)}_{2}+(-1)^{m/2+i_1+i_2}\gamma^{(d)}_1\gamma^{(d)}_2\otimes...\otimes\gamma^{(d)}_1\gamma^{(d)}_2.
\end{equation}
The partial transpositions of the optimal EWs are as follows
\begin{equation}\label{region''}
W^{(m;d;i_1,...,i_{d})T_i}_{ap.,opt.}=I+(-1)^{i_1}\gamma^{(d)}_1\otimes...\otimes
\gamma^{(d)}_1-(-1)^{i_2}\gamma^{(d)}_{2}\otimes...\otimes
\gamma^{(d)}_{2}+(-1)^{m/2+i_1+i_2}\gamma^{(d)}_1\gamma^{(d)}_2\otimes...\otimes\gamma^{(d)}_1\gamma^{(d)}_2.
\end{equation}
Then, the eigenvalues of $W^{(m;d;i_1,...,i_{d})T_i}_{ap.,opt.}$
are given by
\begin{equation}\label{regioneig}
\lambda^{(m;d;i_1,i_{2})}_{k_1,k_{2}}=1+(-1)^{i_1+k_1}-(-1)^{i_2+k_2}+(-1)^{i_1+i_2+k_1+k_2}
\end{equation}
which are not positive with respect to none of the particles. For
example, in the case of $i_1=i_2=0$ the eigenvalues read
\begin{equation}\label{regioneig'}
\lambda^{(m;d;0,0)}_{k_1,k_{2}}=1+(-1)^{k_1}-(-1)^{k_2}+(-1)^{k_1+k_2}
\end{equation}
where, the most negative eigenvalue is $-2$ which is obtained by
taking $k_{1}=1, k_{2}=0$. As before, we consider the density
matrices of the form
\begin{equation}\label{ro}
\rho^{(m;d)}=b_{0}I_{2^{md/2}}+\sum_{k=1}^{d+1}
b_{k}\gamma^{(d)}_k\otimes...\otimes\gamma^{(d)}_k+(-i)^mb_{d+2}\gamma^{(d)}_1\gamma^{(d)}_2\otimes...\otimes\gamma^{(d)}_1\gamma^{(d)}_2
\end{equation}
Then, the positivity of $\rho^{(m;d)}$ implies that
$b_{0}+\sum_{k=1}^{d}(-1)^{i_{k}}b_{k}+(-i)^{md/2}(-1)^{i_1+...+i_{d}}
b_{d+1}+(-i)^m(-1)^{i_1+i_2}b_{d+2}\geq 0$. We discuss the
non-decomposability of $W_{ap.}^{(m;d)}$ only for the case of
$m=2$ and $d=4$, the general cases can be discussed similarly. In
this case, we have
\begin{equation}\label{ro}
\rho^{(2;4)}=\frac{1}{16}I_{16}+\sum_{k=1}^{5}
b_{k}\gamma^{(4)}_k\otimes\gamma^{(4)}_k-b_{6}\gamma^{(4)}_1\gamma^{(4)}_2\otimes\gamma^{(4)}_1\gamma^{(4)}_2
\end{equation}
Then, the vertices of the PPT density matrices' region (the region
defined by the positivity conditions $\rho^{(2;4)}\geq0$ and
$\rho^{(2;4)T_i}\geq0$ for $i=1,2$ which are equivalent to the
inequalities
$\frac{1}{16}+\sum_{k=1}^{4}(-1)^{i_{k}}b_{k}+(-1)^{i_1+...+i_{4}}
b_{5}-(-1)^{i_1+i_2}b_{6}\geq 0$) are given by
\begin{equation}\label{ro4}
\rho^{(2;4)}_{i_1,i_2}=\frac{1}{16}((-1)^{i_1},(-1)^{i_2},0,0,0,-(-1)^{i_1+i_2}),\quad
i_1,i_2\in\{0,1\}.
\end{equation}
In order to determine the region of non-decomposable EWs, we
consider the constraints obtained by
\begin{equation}\label{ineq00'}
Tr(W^{(2;4)}_{ap.}
\rho^{(2;4)}_{i_1,i_2})=16(1-[(-1)^{i_1}a_1+(-1)^{i_2}a_2-(-1)^{i_1+i_2}a_6])<0
\end{equation}
which are equivalent to
$(-1)^{i_1}a_1+(-1)^{i_2}a_2-(-1)^{i_1+i_2}a_6>1$. It can be seen
that the minimum value of
$1-[(-1)^{i_1}a_1+(-1)^{i_2}a_2-(-1)^{i_1+i_2}a_6]$ is obtained by
choosing the parameters $(a_1,...,a_6)$ as
$((-1)^{i_1},(-1)^{i_2},0,0,0,-(-1)^{i_1+i_2})$, which are the
same as the optimal EWs given by (\ref{region'ap}), i.e., all of
the optimal EWs $W^{(2;4;i_1,i_2)}_{ap.,opt.}$ are
non-decomposable. Then, we will have
\begin{equation}\label{ineq00'''}
\mathrm{min}_{_{a_1,a_2,a_6}}(1-[(-1)^{i_1}a_1+(-1)^{i_2}a_2-(-1)^{i_1+i_2}a_6])=-2.
\end{equation}
In fact, the EWs  $W^{(2;4)}_{ap.}$ satisfying the inequalities
(\ref{ineq00'}) are non-decomposable approximate EWs.
\subsection{The second kind}
For the second kind of BSD multispinor EWs we add one of the
multiplications of the matrices $A'_i$, $i=1,2,...,3d/2$, say,
$A'_1A'_2$ to (\ref{EW'}) as
\begin{equation}\label{apWn}
W_{ap.}'^{(m;d)}=a'_{0}I_{2^{md/2}}+\sum_{k=1}^{3d/2}a'_{k}A'_k\otimes...\otimes
A'_k+a'_{3d/2+1}A'_1A'_2\otimes...\otimes A'_1A'_2.
\end{equation}
and try to solve it by LP method. The eigenvalues of
$W_{ap.}'^{(m;d)}$ are given by
\begin{equation}\label{eigap}
a'_{0}+\sum_{k=1}^{d/2+1} (-1)^{i_{k}}
a'_{_{k}}+\sum_{k=2}^{d/2}(-1)^{i_{1}+i_{d/2+1}+
i_{k}}a'_{d/2+k}+\sum_{k=1}^{d/2}(-1)^{m/2+i_{d/2+1}+
i_{k}}a'_{d+k}+(-1)^{i_1+i_2}a'_{3d/2+1},
\end{equation}
for all $i_{1},...,i_{d/2+1}\in \{0,1\}$. The coordinates of the
vertex points which arise from pure product states are listed in
the following table
\begin{equation}\label{tab1ap}
\begin{tabular}{c|c}\hline\hline
Product state &
$(P'_1,...P'_{d/2};P'_{d/2+1},...,P'_{d};P'_{d+1},...,P'_{3d/2};P'_{3d/2+1})$
\\ \hline
$|\psi^{(1;1)}_{\pm}\rangle$ & $(\pm1,1,1,...,1; 0,0,...,0; 0,0,...,0;0)$ \\
$\vdots$ & $\vdots$ \\
$|\psi^{(1;d/2)}_{\pm}\rangle$ & $(1,1,...,1,1,\pm 1;0,0,...,0;0,...,0;0)$ \\
$|\psi^{(2;d/2+1)}_{\pm}\rangle$ & $(0,0,...,0;\pm 1,1,...,1,1;0,0,...,0;0)$\\
$\vdots$ & $\vdots$\\
$|\psi^{(2;d)}_{\pm}\rangle$ & $(0,0,...,0;1,1,...,1,\pm 1;0,0,...,0;0)$\\
$|\psi^{(3;d+1)}_{\pm}\rangle$ & $(0,0,...,0,0;0,...,0,0;\pm 1,1,...,1;0)$\\
$\vdots$ & $\vdots$\\
$|\psi^{(3;3d/2)}_{\pm}\rangle$ & $(0,0,...,0;0,0,...,0;1,...,1,1,\pm 1;0)$\\
$|\psi^{(3;3d/2+1)}_{\pm}\rangle$ & $(0,0,...,0,0;0,0,...,0;0,0,...,0;\pm1)$\\
\hline\hline
\end{tabular}
\end{equation}
where, $\ket{\psi^{(i;k)}_{\pm}}$ are common eigenvectors of the
elements of the commuting set $C_i$ and
$|\psi^{(3;3d/2+1)}_{\pm}\rangle$ is an eigenvector of
$A'_1A'_2\otimes...\otimes A'_1A'_2$.

Now, according to the apexes given by (\ref{tab1ap}), one can
obtain the following inequalities
\begin{equation}\label{ineapprev1}
2+(-1)^{i_1} P'_j+(-1)^{i_2}P'_{j+d/2}+(-1)^{i_3}
P'_{j+d}+(-1)^{i_4}P'_{3d/2+1}\geq 0 ,
\end{equation}
where $i_{1},...,i_{4}\in \{0,1\}$ and $j\in \{1,...,d/2\}$ (the
proof is given in appendix $D$). Therefore, the approximated
feasible region is the intersection of the halfspaces defined by
(\ref{ineapprev1}). The hyperplanes surrounding the feasible
region are given by
\begin{equation}\label{ineapprev1'}
(-1)^{i_1} P'_j+(-1)^{i_2}P'_{j+d/2}+(-1)^{i_3}
P'_{j+d}+(-1)^{i_4}P'_{3d/2+1}= 2.
\end{equation}
The inequalities (\ref{ineapprev1}) imply that the problem does not
lie in the realm of exactly soluble LP problems and we have to use
approximate LP. To this aim we shift aforementioned hyperplanes
parallel to themselves such that they reach to maximum value $2$. On
the other hand the maximum shifting is where the hyperplanes
(\ref{ineapprev1'}) become tangent to convex region coming from pure
product states and in this manner we will be able to encircle the
feasible region by the hyperplanes defined by (\ref{ineapprev1'}).

Regarding the inequalities (\ref{ineapprev1}), the manipulation of
EWs is reduced to the following approximate LP
$$
\hspace{-10cm}\mathrm{minimize} \quad\;\;\;\;\
a'_0+\sum_{i=1}^{3d/2+1}a'_iP'_i\vspace{-1mm}
$$
\begin{equation}\label{region1}
\hspace{0.7cm} \mathrm{subject\ to} \quad \left\{\begin{array}{c}
 2+(-1)^{i_1} P'_j+(-1)^{i_2}P'_{j+d/2}+(-1)^{i_3}
P'_{j+d}+(-1)^{i_4}P'_{3d/2+1}\geq0\\
 \hspace{-8cm}\forall\ |P_k|\leq 1,\\
\end{array}\right.
\end{equation}
for all $i_{1},...,i_{4}\in \{0,1\}$ and $j\in \{1,...,d/2\}$,
where it can be solved by simplex method.

In order that the expectation value of ${W'^{(m;d)}_{ap.}}$ over all
separable states be positive, the following constraints must be
fulfilled
\begin{equation}\label{vpap2'}
|a'_i|\leq \frac{1}{2}a'_0,\quad i=1,...,3d/2+1.
\end{equation}
\subsubsection{The region of non-decomposable (approximate) EWs of the second kind} The region defined by
\begin{equation}\label{ineq1}
a'_{0}+\sum_{k=1}^{d/2+1} (-1)^{i_{k}}
a'_{_{k}}+\sum_{k=2}^{d/2}(-1)^{i_{1}+i_{d/2+1}+
i_{k}}a'_{d/2+k}+\sum_{k=1}^{d/2}(-1)^{m/2+i_{d/2+1}+
i_{k}}a'_{d+k}+(-1)^{i_1+i_2}a'_{3d/2+1}\geq 0,
\end{equation}
for $i_{1},...,i_{d/2+1};i_{3d/2+1}\in\{0,1\}$ is the region where
$W_{ap.}^{'(m;d)}$ is positive.\\
From (\ref{ineapprev1}), it can be seen that the optimal EWs in
the approximated region are given by
$${W'}^{(m;d;i_1,...,i_{d/2+1})}_{ap.,opt.}=I_{2^{md/2}}+\sum_{k=1}^{d/2+1} (-1)^{i_{k}}A'_k\otimes...\otimes
A'_k+\sum_{k=2}^{d/2}(-1)^{i_{1}+i_{d/2+1}+
i_{k}}A'_{d/2+k}\otimes...\otimes A'_{d/2+k}+$$
\begin{equation}\label{region'}
\sum_{k=1}^{d/2}(-1)^{m/2+i_{d/2+1}+
i_{k}}A'_{d+k}\otimes...\otimes
A'_{d+k}+(-1)^{i_1+i_2}A'_1A'_2\otimes...\otimes A'_1A'_2.
\end{equation}
The partial transpositions of optimal EWs
${W'}^{(m;d;i_1,...,i_{d/2+1})}_{ap.,opt.}$ for $m=2,d=4$ are
given by
$${W'}^{(2;4;i_1,i_2,i_3)T_i}_{ap.,opt.}=I_{16}+(-1)^{i_1}A'_1\otimes A'_1+(-1)^{i_2}A'_2\otimes A'_2+(-1)^{i_3}A'_{3}\otimes A'_{3}+(-1)^{i_1+i_2+i_3}A'_4\otimes A'_4+$$
\begin{equation}\label{region'}
(-1)^{i_1+i_3}A'_{5}\otimes A'_{5}+(-1)^{i_2+i_3}A'_6\otimes
A'_6+(-1)^{i_1+i_2}A'_1A'_2\otimes A'_1A'_2.
\end{equation}
The eigenvalues of ${W'}^{(2;4;i_1,i_2,i_{3})T_i}_{ap.,opt.}$ are
given by
$$\lambda'^{(2;4;i_1,i_{2},i_3)}_{k_1,k_{2},k_3}=1+
(-1)^{i_1+k_1}+(-1)^{i_2+k_2}+(-1)^{i_3+k_3}+(-1)^{i_1+i_2+i_3+k_1+k_2+k_3}
-$$
\begin{equation}\label{regioneig}
(-1)^{i_1+i_3+k_1+k_3}-(-1)^{i_2+i_3+k_2+k_3}+(-1)^{i_1+i_2+k_1+k_2}
.
\end{equation}
where, for a given ${W'}^{(2;4;i_1,i_{2},i_3)T_i}_{ap.,opt.}$ are
not necessarily positive for all values of
$k_1,k_2,k_3\in\{0,1\}$, for example for
${W'}^{(2;4;0,0,0)T_i}_{ap.,opt.}$, the most negative eigenvalue
is given by $-4$ which is obtained by taking $k_1=k_2=k_3=1$. This
implies that the optimal EWs
${W'}^{(2;4;i_1,i_{2},i_3)T_i}_{ap.,opt.}$ are not necessarily
decomposable. Now, we consider the density matrices of the form
\begin{equation}\label{ro}
\rho'^{(m;d)}=b'_{0}I_{2^{md/2}}+\sum_{k=1}^{3d/2}b'_{k}A'_k\otimes...\otimes
A'_k+b'_{3d/2+1}A'_1A'_2\otimes...\otimes A'_1A'_2.
\end{equation}
Then, for a bipartite system in the four dimensional space-time,
the vertices of the PPT density matrices' region (the region
defined by the positivity conditions $\rho'^{(2;4)}\geq0$ and
$\rho'^{(2;4)T_i}\geq0$, $i=1,2$), are given by
\begin{equation}\label{ro4ap}
\rho'^{(2;4)}_{i_1,i_2}=\frac{1}{16}((-1)^{i_1},(-1)^{i_2},0,0,0,0,(-1)^{i_1+i_2}),\quad
i_1,i_2\in\{0,1\}.
\end{equation}
Again, by using (\ref{apWn}) and (\ref{ro4ap}), the constraints
obtained by
\begin{equation}\label{ineqap00'}
Tr({\rho'}^{(2;4)}_{i_1,i_2}W_{ap.}^{'(2;4)}
)=16(1-[(-1)^{i_1}a'_1+(-1)^{i_2}a'_2+(-1)^{i_1+i_2}a'_7])<0
\end{equation}
which are equivalent to
$(-1)^{i_1}a'_1+(-1)^{i_2}a'_2-(-1)^{i_1+i_2}a'_7>1$, partially
determine the region of non-decomposable EWs in the approximated
region of EWs. It can be seen that the minimum value of
$1-[(-1)^{i_1}a'_1+(-1)^{i_2}a'_2-(-1)^{i_1+i_2}a'_7]$ is obtained
by choosing the parameters $a'_1,a'_2$ and $a'_7$ as
$(-1)^{i_1},(-1)^{i_2}$ and $-(-1)^{i_1+i_2})$, respectively.
Then, we will have
\begin{equation}\label{ineq00'''}
\mathrm{min}_{_{a'_1,a'_2,a'_7}}(1-[(-1)^{i_1}a'_1+(-1)^{i_2}a'_2-(-1)^{i_1+i_2}a'_7])=-2.
\end{equation}
In fact, the EWs $W_{ap.}^{'(2;4)}$ satisfying the inequalities
(\ref{ineqap00'}) are non-decomposable EWs.
\section{The case of odd $m$}
In this section, we discuss the case of odd number of
$d$-dimensional spinors, briefly. Similar to the case of even $m$,
we need to construct EWs via hermitian commuting operators in
order to calculate the corresponding eigenvalues easily. To do so,
we define two kinds of operators as follows:
\subsection{EWs of the first kind}
In the case of odd $m$, we will consider the following hermitian
matrix
$$W^{(m;d)}=a_0 I_{2^{md/2}}+\sum_{i=1}^{d/2} a_i
\underbrace{\gamma_i^{(d)}\otimes...\otimes
\gamma_i^{(d)}}_{m-1}\otimes A'_i+\sum_{i=1}^{d/2} a_{d/2+i}
\underbrace{\gamma_{d/2+i}^{(d)}\otimes...\otimes
\gamma_{d/2+i}^{(d)}}_{m-1}\otimes A'_i+$$
\begin{equation}\label{Won}
a_{d+1}\underbrace{\gamma_{d+1}^{(d)}\otimes ...\otimes
\gamma_{d+1}^{(d)}}_{m-1}\otimes I_{2^{d/2}},
\end{equation}
where, $A'_i$ for $i=1,2,...,d/2$ are $d/2$ commuting operators
which can be taken from each of three commuting sets $C_1,C_2$ and
$C_3$ defined in (\ref{gamac}).

Again, in order to turn the observable (\ref{Won}) to an EW, we
need to choose the parameters $a_j$, $j=1,2,...,d+1$ in such a way
that it becomes a non-positive operator with positive expectation
values in any pure product state. As in the case of even $m$, in
this case the problem reduces to the LP one, where the feasible
region, EWs' region and the region of detectable entangled states
can be determined similarly.
\subsection{EWs of the second kind}
In the second kind, we consider the following hermitian matrix
$$W'^{(m;d)}=a'_0 I_{2^{md/2}}+\sum_{i=1}^{d/2} a'_i
\underbrace{A'_i\otimes...\otimes A'_i}_{m-1}\otimes
A'_i+\sum_{i=1}^{d/2} a'_{d/2+i}
\underbrace{A'_{d/2+i}\otimes...\otimes A'_{d/2+i}}_{m-1}\otimes
A'_i+$$
\begin{equation}\label{Won'}
\sum_{i=1}^{d/2}a'_{d+i}\underbrace{A'_{d+i}\otimes ...\otimes
A'_{d+i}}_{m-1}\otimes I_{2^{d/2}},
\end{equation}
where, $A'_i$ for $i=1,2,...,d/2$ belong to the commuting set
$C_1$, $A'_{d/2+i}$, $i=1,2,...,d/2$ belong to the commuting set
$C_2$ and $A'_{d+i}$, $i=1,2,...,d/2$ belong to the commuting set
$C_3$.

All of discussions about the second kind of EWs in the case of
even $m$, can be applied in this case similarly.
\section{Conclusion}
Two kinds of Bell-states diagonal multispinor EWs manipulatable
via the exact LP method, were constructed in order to study the
entanglement properties of the relativistic and non-relativistic
multispinor systems in the space-time of arbitrary dimension $d$,
where the first kind can detect some Bell-states diagonal
multispinor PPT entangled states. In particular, in the case of
bipartite system in the four-dimensional space-time, the Bell-type
and iso-concurrence type states were introduced and it was shown
that, these states also the spinor ``EPR" states which are special
kinds of iso-concurrence type entangled states are detected by the
constructed EWs. Moreover, it was shown that the spin entanglement
of a spin entangled BSD density matrix increases under the Lorentz
transformation. The decomposability or non-decomposability of
these EWs was discussed, where the region of non-decomposable EWs
of the first kind was partially determined and the decomposability
of the EWs of the second kind was shown. Also, the EWs for which
the feasible region was not a polygon and the problem was solved
by approximate LP were discussed. Although, we considered only two
kinds of Bell-states diagonal multispinor EWs manipulatable by
exact or approximate LP, it is probable to define some other such
multispinor EWs (even Bell-states non-diagonal multispinor ones)
or some EWs with better approximations (may be solved by exact or
approximate convex optimizations rather than LP ones) such that
the region of PPT entangled states detectable by them be larger,
where all of these cases are under investigation.
\newpage
 \vspace{1cm}\setcounter{section}{0}
 \setcounter{equation}{0}
 \renewcommand{\theequation}{A-\roman{equation}}
  {\Large{Appendix A}}\\
Throughout the paper, we have used the formalism of Euclidean Dirac
fermions, i.e., the analytic continuation to imaginary time
fermionic fields. In this continuation, the pseudo-orthogonal group
$O(d-1,1)$ is replaced with the orthogonal group $O(d)$, $d$ being
the Euclidean space dimension. Therefore Euclidean fermions
transform under the spinorial representation of $O(d)$. In this
appendix we define the algebra of Dirac $\gamma$ matrices and
exhibit matrices which realize the algebra in the Euclidean
representation and explain our notations and conventions.
\\ \textbf{A.1 Dirac $\gamma$ matrices}\\
\textit{Space of even dimensions} $d$. Let $\gamma_{\mu}$,
$\mu=1,...,d$, be a set of $d$ matrices satisfying the anticommuting
relations:
\begin{equation}\label{ap0}
\gamma_{\mu}\gamma_{\nu}+\gamma_{\nu}\gamma_{\mu}=2\delta_{\mu\nu}I,
\end{equation}
in which $I$ is the identity matrix.

These matrices are the generatores of a Clifford algebra similar to
the algebra of operators acting on Grassmann algebras. It follows
from relations (\ref{ap0}) that the $\gamma$ matrices generate an
algebra which, as a vector space, has a dimension $2^{d}$. In the
following, we will give an inductive construction $(d\rightarrow
d+2$) of hermitian matrices satisfying (\ref{ap0}). In the algebra
one element plays a special role, the product of all $\gamma$
matrices. The matrix $\gamma_S$:
\begin{equation}\label{ap1}
\gamma_{S}=i^{-d/2}\gamma_1\gamma_2...\gamma_{2n},
\end{equation}
anticommutes, because $d$ is even, with all other $\gamma$ matrices
and $\gamma^2_{S}=I$.

In calculations involving $\gamma$ matrices, it is not always
necessary to distinguish $\gamma_S$ from other $\gamma$ matrices.
Identifying thus $\gamma_S$ with $\gamma_{d+1}$, we have:
\begin{equation}\label{ap3}
\gamma_{i}\gamma_{j}+\gamma_{j}\gamma_{i}=2\delta_{ij}I,\;\;\
i,j=1,...,d,d+1.
\end{equation}
The Greek letters $\mu$ $\nu...$ are usually used to indicate that
the value $d+1$ for the index has been excluded.\\
\textit{Space of odd dimensions}. Equation (\ref{ap3}) shows that in
odd dimensions, we can represent the $\gamma$ matrices by taking the
$\gamma$ matrices of dimension $d-1$, to which we add $\gamma_S$.
Note, however that in this case, in contrast to the even case, the
$\gamma$ matrices are not
all algebraically independent.\\
\textbf{A.2 An explicit construction of $\gamma^{(d)}_i$} \\
It is sometimes useful to have an explicit realization of the
algebra of $\gamma$ matrices.

For $d=2$, the standard Pauli matrices realize the algebra:
$$\gamma^{(d=2)}_{1}\equiv \sigma_1=\left(\begin{array}{cc}
           0 & 1 \\
           1 & 0 \\
         \end{array}\right),\;\ \gamma^{(d=2)}_{2}\equiv \sigma_2=\left(\begin{array}{cc}
           0 & -i \\
           i & 0 \\
         \end{array}\right),$$
\begin{equation}\label{ap4}
\gamma^{(d=2)}_{S}\equiv
\gamma^{(d=2)}_{3}\equiv\sigma_3=\left(\begin{array}{cc}
           1 & 0 \\
           0 & -1 \\
         \end{array}\right).
\end{equation}
The three matrices are hermitian, i.e., $\gamma_i=\gamma^{\dag}_i$.
The matrices $\gamma_1$ and $\gamma_3$ are symmetric and $\gamma_2$
is antisymmetric, i.e., $\gamma_1=\gamma^t_1$, $\gamma_3=\gamma^t_3$
and $\gamma_2=-\gamma^t_2$.

To construct the $\gamma$ matrices for higher even dimensions, we
then proceed by induction, setting:
$$\gamma^{(d+2)}_i=\sigma_1\otimes \gamma^{(d)}_i=\left(\begin{array}{cc}
                                      0 & \gamma^{(d)}_i \\
                                                \gamma^{(d)}_i & 0 \\
                                              \end{array}\right),\;\
i=1,...,d+1,$$
\begin{equation}\label{ap4}
\gamma_{d+2}=\sigma_2\otimes I^{(d)}=\left(\begin{array}{cc}
                                      0 & -iI_d \\
                                                iI_d & 0 \\
                                              \end{array}\right),
\end{equation}
where, $I_d$ is the unit matrix in $2^{d/2}$ dimensions.

As a consequence $\gamma^{(d+2)}_S$ has the form:
\begin{equation}\label{ap4'}
\gamma^{(d+2)}_S\equiv\gamma^{(d+2)}_{d+3}=\sigma_3\otimes
I_d=\left(\begin{array}{cc}
                                      I_d & 0 \\
                                                0 & -I_d \\
                                              \end{array}\right).
\end{equation}
A straightforward calculation shows that if the matrices
$\gamma_i^{(d)}$ satisfy relations (\ref{ap3}), the
$\gamma_i^{(d+2)}$ matrices satisfy the same relations. By induction
we see that the $\gamma$ matrices are all hermitian. from
(\ref{ap4}), it is seen that, if $\gamma_i^{(d)}$ is symmetric or
antisymmetric, $\gamma_i^{(d+2)}$ has the same property. The matrix
$\gamma_{d+2}^{(d+2)}$ is antisymmetric and $\gamma_S^{(d+2)}$ which
is also $\gamma_{d+3}^{(d+2)}$ is symmetric. It follows immediately
that, in this representation, all $\gamma$ matrices with odd index
are symmetric and all matrices with even index are antisymmetric,
i.e.,
\begin{equation}\label{ap4'}
\gamma^t_i=(-1)^{i+1}\gamma_i.
\end{equation}
{\Large{Appendix B}}\\
In this appendix we  prove  the inequalities
(\ref{ine0'}) and (\ref{Cn}).\\
\textbf{Proof of the inequalities (\ref{ine0'}):}\\
In order to prove the inequalities (\ref{ine0'}), we first prove
that the expectation value of the operator
$I+\sum_{k=1}^{d+1}(-1)^{i_k}\underbrace{\gamma_k^{(d)}\otimes...\otimes
\gamma_k^{(d)}}_m$ over an arbitrary pure product state
$\ket{\alpha_1}\ket{\alpha_2}...\ket{\alpha_m}$ is non-negative.

By defining
$b_i:=\langle\psi^{(d)}|\gamma_i^{(d)}|\psi^{(d)}\rangle$, where
$|\psi^{(d)}\rangle$ is an arbitrary pure state in the Hilbert space
of dimension $2^{d/2}$, first we prove that
$\Sigma_{i=1}^{2d+1}b^2_i\leq 1$. We prove this by induction on $d$.
First note that by using (\ref{ap4}), the matrices $\gamma_i^{(d)}$
can be rewritten recursively as follows
\begin{equation}\label{Wn1}
\gamma_1^{(d)}=\gamma_1^{(d-2)}\otimes \sigma_1,\;\
\gamma_2^{(d)}=\gamma_1^{(d-2)}\otimes \sigma_2,\;\
\gamma_3^{(d)}=\gamma_1^{(d-2)}\otimes \sigma_3,\;\
\gamma_i^{(d)}=\gamma_{i-2}^{(d-2)}\otimes I_2 ,\;\ i=4,...,d+1.
\end{equation}

Now, we consider the pure state $\ket{\psi^{(d)}}$ as follows
\begin{equation}\label{ap'}
\ket{\psi^{(d)}}=\beta_d\ket{\psi^{(d-2)}}\ket{+x}+\delta_d\ket{\psi'^{(d-2)}}\ket{-x},\;\
|\beta_d|^2+|\delta_d|^2=1.
\end{equation}
By using (\ref{Wn1}), it is seen that by a rotation of magnitude
$\pi/2$ about the $x$ axes in the last component of
$\gamma_i^{(d)}$, one can take the expectation values of
$\gamma_2^{(d)}$ and $\gamma_3^{(d)}$ equal to zero, i.e.,
$b_2=b_3=0$ (recall that
$\gamma_1^{(d)}=\sigma_1\otimes...\otimes\sigma_1$,
$\gamma_2^{(d)}=\sigma_1\otimes...\otimes\sigma_1\otimes \sigma_2$
and
$\gamma_3^{(d)}=\sigma_1\otimes...\otimes\sigma_1\otimes\sigma_3$).
Therefore, we have
$$b_1=\langle\psi^{(d)}|\gamma_1^{(d)}|\psi^{(d)}\rangle=|\alpha_d|^2\langle\psi^{(d-2)}|\gamma_1^{(d-2)}|\psi^{(d-2)}\rangle-|\beta_d|^2\langle\psi'^{(d-2)}|\gamma_1^{(d-2)}|\psi'^{(d-2)}\rangle,$$
\begin{equation}\label{ap'1}
b_i=\langle\psi^{(d)}|\gamma_i^{(d)}|\psi^{(d)}\rangle=|\alpha_d|^2\langle\psi^{(d-2)}|\gamma_{i-2}^{(d-2)}|\psi^{(d-2)}\rangle+|\beta_d|^2\langle\psi'^{(d-2)}|\gamma_{i-2}^{(d-2)}|\psi'^{(d-2)}\rangle,\;\
i=4,...,2d+1.
\end{equation}
Then, we have
$$\sum_ib_i^2=|\alpha_d|^4\underbrace{(\langle\psi^{(d-2)}|\gamma_1^{(d-2)}|\psi^{(d-2)}\rangle^2+\sum_{i=4}^{2d+1}\langle\psi^{(d-2)}|\gamma_{i-2}^{(d-2)}|\psi^{(d-2)}\rangle^2)}_{\leq1}+$$
$$|\beta_d|^4\underbrace{(\langle\psi'^{(d-2)}|\gamma_1^{(d-2)}|\psi'^{(d-2)}\rangle^2
+\sum_{i=4}^{2d+1}\langle\psi'^{(d-2)}|\gamma_{i-2}^{(d-2)}|\psi'^{(d-2)}\rangle^2)}_{\leq1}+
2|\alpha_d|^2|\beta_d|^2\{\sum_{i=4}^{2d+1}\langle\psi^{(d-2)}|\gamma_{i-2}^{(d-2)}|\psi^{(d-2)}\rangle\times$$
$$\langle\psi'^{(d-2)}|\gamma_{i-2}^{(d-2)}|\psi'^{(d-2)}\rangle-
\langle\psi^{(d-2)}|\gamma_1^{(d-2)}|\psi^{(d-2)}\rangle\langle\psi'^{(d-2)}|\gamma_1^{(d-2)}|\psi'^{(d-2)}\rangle\}\leq$$
$$|\alpha_d|^4+|\beta_d|^4+2|\alpha_d|^2|\beta_d|^2\sqrt{\underbrace{\sum_{i=3}^{2d+1}\langle\psi^{(d-2)}|\gamma_{i-2}^{(d-2)}|\psi^{(d-2)}\rangle^2}_{\leq1}.\underbrace{\sum_{i=3}^{2d+1}\langle\psi'^{(d-2)}|\gamma_{i-2}^{(d-2)}|\psi'^{(d-2)}\rangle^2}_{\leq1}}\leq$$
\begin{equation}\label{ap'2}
|\alpha_d|^4+|\beta_d|^4+2|\alpha_d|^2|\beta_d|^2=1,
\end{equation}
where, we have used the hypothesis of induction in the first two
inequalities and the Schwartz inequality  in the third one.

Now, by using the fact that
$|\langle\alpha_i|\gamma_k^{(d)}|\alpha_i\rangle|\leq 1$,
$i=1,2,...,m$ we have
$$
\hspace{-1.5cm}Tr(\sum_{k=1}^{d+1}(-1)^{i_k}\underbrace{\gamma_k^{(d)}\otimes...\otimes
\gamma_k^{(d)}}_m|\alpha_1\rangle\langle\alpha_1|\otimes...\otimes
|\alpha_m\rangle\langle\alpha_m|)\leq$$$$\sum_{k=1}^{d+1}|\langle\alpha_1|\gamma_k^{(d)}|\alpha_1\rangle\langle\alpha_2|\gamma_k^{(d)}|\alpha_2\rangle...\langle\alpha_m|\gamma_k^{(d)}|\alpha_m\rangle|\leq$$
\begin{equation}\label{eqq}
\sum_{k=1}^{d+1}|\langle\alpha_1|\gamma_k^{(d)}|\alpha_1\rangle\langle\alpha_2|\gamma_k^{(d)}|\alpha_2\rangle|\leq\sqrt{\underbrace{\sum_{k=1}^{d+1}(\langle\alpha_1|\gamma_k^{(d)}|\alpha_1\rangle)^2}_{\leq1}.\underbrace{\sum_{k=1}^{d+1}\langle\alpha_2|\gamma_k^{(d)}|\alpha_2\rangle)^2}_{\leq
1}}\leq1,
\end{equation}
where, we have used the Schwartz inequality in the third inequality
and the fact that
$\sum_{i=1}^{d+1}b^2_i=\sum_{i=1}^{d+1}(\langle\psi^{(d)}|\gamma_i^{(d)}|\psi^{(d)}\rangle)^2\leq1$.

 Therefore, the expectation value of the operator $I+\sum_{k=1}^{d+1}(-1)^{i_k}\underbrace{\gamma_k^{(d)}\otimes...\otimes
\gamma_k^{(d)}}_m$ over any pure product state is non-negative,
hence it is non-negative over any separable state $\rho_s$, since
separable states can be written as convex combinations of pure
product states.\\
\textbf{Proof of the inequalities (\ref{Cn}):}\\ We consider the
case $j=1$; $i_1=i_2=i_3=0$, the proof of the other cases is
similar. As regards the arguments of the proof of inequalities
(\ref{ine0'}), it must be proved that the expectation value of the
operator $I+A'_1\otimes...\otimes A'_1+A'_{d/2+1}\otimes...\otimes
A'_{d/2+1}+ A'_{d+1}\otimes...\otimes A'_{d+1}$ over the pure
product state $\ket{\alpha_1}...\ket{\alpha_m}$ is non-negative.

Now, by using the fact that
$|\langle\alpha_i|A'_k|\alpha_i\rangle|\leq 1$, $i=1,2,...,m$ we
have
$$
\hspace{-0.5cm}Tr\{(A'_1\otimes...\otimes
A'_1+A'_{d/2+1}\otimes...\otimes A'_{d/2+1}+
A'_{d+1}\otimes...\otimes
A'_{d+1})|\alpha_1\rangle\langle\alpha_1|\otimes...\otimes
|\alpha_m\rangle\langle\alpha_m|\}\leq$$$$|\langle\alpha_1|A'_1|\alpha_1\rangle...\langle\alpha_m|A'_{1}|\alpha_m\rangle|+|\langle\alpha_1|A'_{d/2+1}|\alpha_1\rangle...\langle\alpha_m|A'_{d/2+1}|\alpha_m\rangle|+|\langle\alpha_1|A'_{d+1}|\alpha_1\rangle...\langle\alpha_m|A'_{d+1}|\alpha_m\rangle|\leq$$
\begin{equation}\label{eqq}
|\langle\alpha_1|A'_1|\alpha_1\rangle|+|\langle\alpha_1|A'_{d/2+1}|\alpha_1\rangle|+|\langle\alpha_1|A'_{d+1}|\alpha_1\rangle|\leq
1,
\end{equation}
where, we have used the fact that
$A'_1=-i\gamma^{(d)}_1\gamma^{(d)}_2=I\otimes...\otimes I\otimes
\sigma_z$, $A'_{d/2+1}=\gamma^{(d)}_1=\sigma_x\otimes...\otimes
\sigma_x$ and $A'_{d+1}=\gamma^{(d)}_2=\sigma_x\otimes...\otimes
\sigma_x\otimes \sigma_y$ and so, by a rotation of magnitude $\pi/2$
about the $z$ axis in the last component of $A'_1$, $A'_{d/2+1}$ and
$A'_{d+1}$, one can take the expectation values
$|\langle\alpha_1|A'_{d/2+1}|\alpha_1\rangle|$ and
$|\langle\alpha_1|A'_{d+1}|\alpha_1\rangle|$ equal to zero. $\Box$\\
{\Large{Appendix C}}\\
In this appendix, we show that the region of SSNNEV is convex if
the feasible region be convex.

Let $W=a_0 I+\sum_ia_iO_i$ be a hermitian operator. Then, in order
that $W$ be an EW, the function $F(a,P)$ defined as
\begin{equation}\label{apc0}
F(a,P)=a^TP+a_0
\end{equation}
must be positive ($P_i:=Tr(O_i\rho_s)$ for any separable state
$\rho_s$), hence, the region of SSNNEV is defined by
\begin{equation}\label{apc1}
\mathrm{inf}_{_{P}} F(a,P)=\mathrm{inf}_{_{P}}(a^TP+a_0)\geq 0.
\end{equation}
Now, it must be proved that the region defined by (\ref{apc1}) is
convex. To do so, note that $F(a,P)$ is an affine and therefore also
linear function (recall that a function is affine if it is a sum of
a linear function and a constant). Then, it is
both convex and concave \cite{Boyd}. Now, we recall the definition of the conjugate function and sublevel sets of a function as follows:\\
\textbf{Definition 1} Let $f:R^n\rightarrow R$. The function
$f^*:R^n\rightarrow
 R$ defined as
\begin{equation}\label{apc2}
f^*(y)=sup_{_{x\in \mathrm{dom}  f}}(y^Tx-f(x)),
\end{equation}
is called the conjugate of the function $f$ ($\mathrm{dom}$ denotes
the domain of $f$).

It is seen immediately that $f^*$ is a convex function, since it is
the pointwise supremum of a family of convex (indeed, affine)
functions of $y$. This is true whether or not $f$ is convex.\\
\textbf{Definition 2} The $\alpha$-sublevel set of a function
$f:R^n\rightarrow R$ is defined as
\begin{equation}\label{apc3}
C_{\alpha}=\{\alpha\in \mathrm{dom}  f | f(x)\leq \alpha\}.
\end{equation}
Sublevel sets of a convex function are convex, for any value of
$\alpha$ \cite{Boyd}.

Now, we consider the conjugate function of the constant function
$f(P)=a_0$, for all $P$ in the feasible region. Then, (\ref{apc1})
is equivalent to
\begin{equation}\label{apc3}
\mathrm{sup}_{_{P}}(-a^TP-a_0)\leq0.
\end{equation}
By renaming $P'=-P$, (\ref{apc3}) is written as
\begin{equation}\label{apc4}
f^*(a)=\mathrm{sup}_{_{P'}}(a^TP'-a_0)\leq0.
\end{equation}
It could be noticed that, the set $\{a\in \mathrm{dom} f^* |
f^*(a)\leq 0\}$ is the $0$-sublevel set of the convex function $f^*$
and so is a convex set. Therefore, we conclude that the set $\{a\in
\mathrm{dom} f^* | \mathrm{inf}_{_{P}}(a^TP+a_0)\geq 0 \}$ is
convex. $\Box$

It should be noticed that if the feasible region be a polygon,
then the region of SSNNEV is also a polygon. Therefore, the apexes
of the feasible region correspond to the hyperplanes surrounding
the region of SSNNEV and vice versa, i.e., the feasible region and
the region of
SSNNEV are dual with each other. \\
{\Large{Appendix D}}\\
{\bf Proof of the inequalities (\ref{ineapprev})}:\\
We prove the Eq.(\ref{ineapprev}) only for the case
$i_1=...=i_{d+2}=0$. The proof of the other cases is similar. Then,
the Eq.(\ref{ineapprev}) is given by
\begin{equation}\label{apd}
\sqrt{2}+\sum_{k=1}^{d+2}P_{k}\geq 0.
\end{equation}
As before, it is sufficient to prove that the expectation value of
the operator $\sqrt{2}I+\sum_{k=1}^{d+2}A_k\otimes...\otimes A_k$
with $A_{d+2}= \gamma^{(d)}_1\gamma^{(d)}_2$, over any pure product
state $|\alpha_1\rangle...|\alpha_m\rangle$ is non-negative
($\sum_{k=1}^{d+2}P_{k}$ is the expectation value of the operator
$\sum_{k=1}^{d+2}A_k\otimes...\otimes A_k$, over any separable
state). To do so, we define
$b_i=\langle\alpha_1|A_i|\alpha_1\rangle$, for $i=1,...,d+2$ and
evaluate the largest eigenvalue of $\sum_{k=1}^{d+2}b_kA_k$. Now, we
note that
\begin{equation}\label{apd2}
(\sum_{k=1}^{d+2}b_kA_k)^2=(\sum_{k=1}^{d+1}b^2_k+b^2_{d+2})I,
\end{equation}
where, we have used the fact that $A_{d+2}$ anticommutes with $A_1$
and $A_2$. Therefore, the eigenvalues of $(b_1A_1+b_2A_2\pm
b_{d+2}A_{d+2})^2$ are given by
\begin{equation}\label{apd3}
\lambda^2=b^2_1+b^2_2+b^2_{d+2}\leq1+\cos^22\theta \leq 2,
\end{equation}
where, we have used the fact that
\begin{equation}\label{apd4}
b_{d+2}=\langle\alpha|A_{d+2}|\alpha\rangle=\langle\alpha|I\otimes...\otimes
I\otimes
\sigma_z|\alpha\rangle=\sum_{k=1}^{2^{d/2-1}}|\alpha_{2k-1}|^2-\sum_{k=1}^{2^{d/2-1}}|\alpha_{2k}|^2=1-2\sum_{k=1}^{2^{d/2-1}}|\alpha_{2k}|^2.
\end{equation}
From the equality $\sum_{k=1}^{2^{d/2}}|\alpha_{k}|^2=1$, it can be
seen that one can choose a parametrization for $\alpha_i$ such that
$\sum_{k=1}^{2^{d/2-1}}|\alpha_{2k-1}|^2=\cos^2\theta$ and
$\sum_{k=1}^{2^{d/2-1}}|\alpha_{2k}|^2=\sin^2\theta$. Then,
(\ref{apd4}) will imply that
$b_{d+2}=1-2\sin^2\theta=\cos2\theta$.\\
{\bf Proof of the  inequalities (\ref{ineapprev1})}:\\
We consider only the case of $i_1=i_2=i_3=0$ and $j=1$. Then, the
Eq.(\ref{ineapprev1}) is given by
\begin{equation}\label{apd5}
P'_{1}+P'_{d/2+1}-(-i)^mP'_{d+1}+P'_{3d/2+1}=P'_{1}+P'_{d/2+1}\pm
P'_{d+1}+P'_{3d/2+1}\leq 2 ,
\end{equation}
Now, similar to the proof of Eq.(\ref{ineapprev}) as in the above,
we prove that the expectation value of the operator $2I+A'_1\otimes
A'_1+A'_{d/2+1}\otimes A'_{d/2+1}\pm A'_{d+1}\otimes
A'_{d+1}+A'_{3d/2+1}\otimes A'_{3d/2+1}$ over any pure product state
$|\alpha_1\rangle|\alpha_2\rangle$ is non-negative. By defining
$b'_i=\langle\alpha_1|A'_i|\alpha_1\rangle$, for
$i=1,\frac{d}{2}+1,d+1,\frac{3d}{2}+1$, we need to evaluate the
largest eigenvalue of $b'_1A'_1+b'_{d/2+1}A'_{d/2+1}\pm
b'_{d+1}A'_{d+1}+b'_{3d/2+1}A'_{3d/2+1}$ as before. One can easily
check that
\begin{equation}\label{apd7}
(b'_1A'_1+b'_{d/2+1}A'_{d/2+1}\pm
b'_{d+1}A'_{d+1}+b'_{3d/2+1}A'_{3d/2+1})^2=\sum_{i}b'^2_iI+2b'_1b'_{\frac{3d}{2}+1}A'_1A_{\frac{3d}{2}+1},
\end{equation}
where, we have used the fact that $A'_{\frac{3d}{2}+1}$ anticommutes
with $A'_{d/2+1}$ and $A_{d+1}$ and commutes with $A'_1$. Then, the
eigenvalues of $(b'_1A'_1+b'_{d/2+1}A'_{d/2+1}\pm
b'_{d+1}A'_{d+1}+b'_{3d/2+1}A'_{3d/2+1})^2$ are as follows
$$\lambda'=\sum_{i}{b'}_i^2\pm2b'_1b'_{\frac{3d}{2}+1}\leq
1+b'_{\frac{3d}{2}+1}(b'_{\frac{3d}{2}+1}+2b'_1)= 1+\sin
2\phi(\cos^2\theta\cos\theta'-\sin^2\theta\cos\theta'')\times$$
\begin{equation}\label{apd8}
[2\cos2\theta+\sin2\phi(\cos^2\theta\cos\theta'-\sin^2\theta\cos\theta'')]\leq
4,
\end{equation}
where, the maximum value $4$ is obtained by taking $\phi=\pi/4,
\theta=\theta'=0$. Note that above, we have used the following
equality
\begin{equation}\label{apd9}
b'_{3d/2+1}=\langle\alpha|\gamma^{(d)}_3|\alpha\rangle=\langle\alpha|\sigma_x\otimes...\otimes
\sigma_x \otimes \sigma_z|\alpha\rangle=2\{
Re(\sum_{k=1}^{2^{d/2-2}}\alpha^*_{2k-1}\alpha_{2^{d/2}-2k+1})-Re(\sum_{k=1}^{2^{d/2-2}}\alpha^*_{2k}\alpha_{2^{d/2}-2k})\}.
\end{equation}
{\Large{Appendix E}}\\
{\bf Proof of the inequalities (\ref{BD0})}:\\
First we note that, by applying the transform $H\otimes I$ with
$H=\frac{1}{\sqrt{2}}(\sigma_x+\sigma_z)$ on the first particle, the
helicity basis (\ref{helicity}) take the following form
\begin{equation}\label{bell00}
|\psi_1\rangle=\ket{00},\;\ |\psi_2\rangle=\ket{11}\;\
|\psi_3\rangle=\ket{10}\;\ |\psi_4\rangle=\ket{01}
\end{equation}
which are the same as Dirac's spinors. Also, this transformation
changes the Bell-type states $\ket{\Psi_i}$, $i=1,2,...,16$ to the
traditional Bell states \cite{akhtar}-\cite{rez} which are
obtained via the action of the Heisenberg group $H_{Z_2\times
Z_2}(\cong (Z_2\times Z_2)\times(Z_2\times Z_2)\rtimes (Z_2\times
Z_2)$) on the following maximally entangled state
\begin{equation}\label{MES}
\ket{\Psi_{00}}=\frac{1}{2}\sum_{i,j=0,1}\ket{ij}\ket{ij},
\end{equation}
i.e., we have
\begin{equation}\label{MES1}
\ket{\Psi_{\mu\nu}}=A_{\mu}\otimes
A_\nu\ket{\Psi_{00}}=\sigma_\alpha\otimes \sigma_\beta\otimes
\sigma_{\alpha'}\otimes
\sigma_{\beta'}=\Omega^iS^j\otimes\Omega^kS^l\otimes\Omega^{i'}S^{j'}\otimes\Omega^{k'}S^{l'}\ket{\Psi_{00}},
\end{equation}
where the operators $S=\sigma_x$ and $\Omega=\sigma_z$ known as
shift and modulation operators are the generators of the Heisenberg
group $H_{Z_2\otimes Z_2}$.
 Then, it is sufficient to show that
$\ket{\Psi_{00}}\langle\Psi_{00}|$ is written in terms of the
diagonal elements $A_{\mu}\otimes A_\mu$. To do so, let
\begin{equation}\label{MES0}
\ket{\Psi_{00}}\langle\Psi_{00}|=\sum_{\mu,\nu}b_{\mu\nu}A_{\mu}\otimes
A_\nu.
\end{equation}
where, $ b_{\mu\nu}=\langle\Psi_{00}|A_{\mu}\otimes
A_\nu|\Psi_{00}\rangle$. By taking $A_{\mu}=\sigma_\alpha\otimes
\sigma_\beta$ and $A_{\nu}=\sigma^{\dag}_{\alpha'}\otimes
\sigma^{\dag}_{\beta'}$ and using (\ref{MES}), we obtain
$$
b_{\mu\nu}=\sum_{i,i',j,j'}\langle i|\sigma_\alpha|i'\rangle\langle
i|\sigma^{\dag}_{\alpha'}|i'\rangle\langle
j|\sigma_\beta|j'\rangle\langle
j|\sigma^{\dag}_\beta|j'\rangle=\sum_{i,i',j,j'}\langle
i|\Omega^kS^l|i'\rangle\langle
i|S^{-l'}\Omega^{-k'}|i'\rangle\langle
j|\Omega^rS^s|j'\rangle\langle j|S^{-s'}\Omega^{-r'}|j'\rangle$$
$$=\sum_{i,i',j,j'}\langle i|\Omega^k|l+i'\rangle\langle
i-l'|\Omega^{-k'}|i'\rangle\langle j|\Omega^r|s+j'\rangle\langle
j-s'|\Omega^{-r'}|j'\rangle=\sum_{i,i',j,j'}\omega^{(k-k')i}\omega^{(r-r')j}\delta_{i,l+i'}\delta_{i-l',i'}\delta_{j,s+j'}\delta_{j-s',j'}
$$
\begin{equation}\label{MES000}
=\delta_{kk'}\delta_{ll'}\delta_{rr'}\delta_{ss'}=\delta_{\alpha,\alpha'}\delta_{\beta,\beta'}=\delta_{\mu,\nu}\quad,\quad
\omega=e^{-\pi i}=-1.
\end{equation}\\
\textbf{Proof for the fact that $W_{opt}(p)$ given in (\ref{Wp}) is
an entanglement witness}\\
In order to show that $W_{opt}(p)$ in (\ref{Wp}) is an EW, it must
be proved that the expectation value of $W_{opt}(p)$ over any
product state $\ket{\gamma}=\ket{\alpha}\ket{\beta}$ is
non-negative. To do so, as it is seen from Eq.(\ref{HSm1}), we need
to show that
\begin{equation}\label{ape}
\langle\gamma|\rho_s(p)-\rho^{(1,0,0,0)}_{_{ent}}(p)|\gamma\rangle-\varepsilon(p)\geq0.
\end{equation}
In order to prove (\ref{ape}), first we evaluate the minimum value
of
$\langle\gamma|\rho_s(0)-\rho^{(1,0,0,0)}_{_{ent}}(0)|\gamma\rangle$
as follows

$$\langle\gamma|\rho_s(0)-\rho^{(1,0,0,0)}_{_{ent}}(0)|\gamma\rangle=\frac{1}{120}\{\langle\gamma|\gamma^0\otimes\gamma^0|\gamma\rangle-\langle\gamma|\gamma^1\otimes\gamma^1|\gamma\rangle-\langle\gamma|\gamma^2\otimes\gamma^2|\gamma\rangle-\langle\gamma|\gamma^3\otimes\gamma^3|\gamma\rangle-\langle\gamma|\gamma^5\otimes\gamma^5|\gamma\rangle\}=$$
\begin{equation}\label{ape1}
\frac{1}{120}\{b_0\langle\beta|\gamma^0|\beta\rangle-b_1\langle\beta|\gamma^1|\beta\rangle-b_2\langle\beta|\gamma^2|\beta\rangle-b_3\langle\beta|\gamma^3|\beta\rangle-b_5\langle\beta|\gamma^5|\beta\rangle\},
\end{equation}
with $b_{\mu}:=\langle\alpha|\gamma^{\mu}|\alpha\rangle$ for
$\mu=0,1,2,3,5$. By defining
\begin{equation}\label{ape2}
O:=b_0\gamma^0-b_1\gamma^1-b_2\gamma^2-b_3\gamma^3-b_5\gamma^5,
\end{equation}
and using the fact that the eigenvalues of $O$ are
$\pm\sqrt{b_0^2+b_1^2+b_2^2+b_3^2+b_5^2}$ (from the
anti-commutativity of $\gamma^{\mu}$, $\mu=0,1,2,3,5$ we have
$O^2=(b_0^2+b_1^2+b_2^2+b_3^2+b_5^2)I\otimes I$), we obtain
\begin{equation}\label{ape3}
\langle\gamma|\rho_s(0)-\rho^{(1,0,0,0)}_{_{ent}}(0)|\gamma\rangle=\frac{1}{120}\langle\beta|O|\beta\rangle\geq
-\frac{1}{120},
\end{equation}
where, we have used the fact that
$\sqrt{b_0^2+b_1^2+b_2^2+b_3^2+b_5^2}\leq 1$ (see the proof of the
inequalities (\ref{ine0'}) given in appendix $B$). Therefore the
minimum value of
$\langle\gamma|\rho_s(0)-\rho^{(1,0,0,0)}_{_{ent}}(0)|\gamma\rangle$
is equal to $-\frac{1}{120}$. Then, we can write
$$
\langle\gamma|\rho_s(p)-\rho^{(1,0,0,0)}_{_{ent}}(p)|\gamma\rangle=\frac{1}{\cosh^2(\xi)}\langle\gamma|(D\otimes
D)(\rho_s(0)-\rho^{(1,0,0,0)}_{_{ent}}(0))(D^{\dag}\otimes
D^{\dag})|\gamma\rangle=$$
\begin{equation}\label{ape4}
\frac{1}{\cosh^2(\xi)}\langle\gamma'|(\rho_s(0)-\rho^{(1,0,0,0)}_{_{ent}}(0))|\gamma'\rangle\geq
-\frac{1}{120\cosh^2(\xi)},
\end{equation}
where, $\ket{\gamma'}:=(D^{\dag}\otimes D^{\dag})|\gamma\rangle$ is
another product state and so the expectation value of
$\rho_s(0)-\rho^{(1,0,0,0)}_{_{ent}}(0)$ over it is larger than
$-\frac{1}{120}$. Therefore, by using (\ref{epsilonp}) and
(\ref{ape4}), one can obtain
$$\langle\gamma|\rho_s(p)-\rho^{(1,0,0,0)}_{_{ent}}(p)|\gamma\rangle-\varepsilon(p)\geq\frac{1}{120\cosh^2(\xi)}\{\frac{\cosh^8(\xi/2)}{5\cosh^4(\xi)}[5(1+\tanh^8(\xi/2))+$$
\begin{equation}\label{ape5}
28(\tanh^2(\xi/2)+\tanh^6(\xi/2))+126\tanh^4(\xi/2)]-1\}\geq0.
\end{equation}


\begin{thebibliography}{99}
\bibitem{lew} M. Lewenstein, D. Bruss, J. I. Cirac, M. Ku´s, J.
Samsonowicz, A. Sanpera and R. Tarrach, J. Mod. Opt. 77, 2481
(2000).
\bibitem{4'}
B. M. Terhal, Phys. Lett. A 271, 319 (2000).
\bibitem{11}
B. M. Terhal, Theor. Comput. Sci. 287(1), 313 (2002).
\bibitem{13}
A. C. Doherty, P. A. Parrilo, and F. M. Spedalieri, Phys. Rev. A
69, 022308 (2004).
\bibitem{14}
M. Lewenstein, B. Kraus, J. I. Cirac, and P. Horodecki, Phys. Rev.
A 62, 052310 (2000).
\bibitem{5}
A. Jamiolkowski, Rep. Math. Phys. 3, 275 (1972).
\bibitem{15}
M. Lewenstein, B. Kraus, P. Horodecki, and J. I. Cirac, Phys. Rev.
A 63, 044304 (2001).
\bibitem{akhtar}
S. J. Akhtarshenas and M.A. Jafarizadeh, The European Physical
Journal D Vol. 25 No.3, 293 (2003).
\bibitem{akhtar3}
S. J. Akhtarshenas and M.A. Jafarizadeh, J. Phys. A : Math. Gen.
37, 2965 (2004).
\bibitem{akhtar2}
S. J. Akhtarshenas and M.A. Jafarizadeh, Quantum Information and
computation, Vol. 3, No. 3, 229 (2003).
\bibitem{rez}
 M. A. Jafarizadeh, M. Mirzaee and M. Rezaee, International Journal of Quantum Information (IJQI) Vol.3, No.
 3, 511 (2005).
\bibitem{12}
D. Bruss et al., J. Mod. Opt. 49, 1399 (2002).
\bibitem{brus}
K. Eckert, J. Schliemann, D. Bruss and M. Lewenstein, Annals of
physics, 299, 88 (2002).
\bibitem{2}
M. N. Nielsen and I. L. Chuang, Quantum Computation and Quantum
Information (Cambridge University Press, Cambridge, England,
2000).
\bibitem{telep}
C.H. Bennett, G. Brassard, C. Crepeau, R. Jozsa, A. Peres, and
W.K. Wooters, Phys. Rev. Lett. 69, 2881 (1992).
\bibitem{cync}
W. Y. Hwang, D. Ahn, S. W. Hwang and Y. D. Han, Eur. Phys. J. D19,
129 (2002).
\bibitem{ryder}
L. H. Ryder, Quantum Field Theory (Cambridge University Press, New
York, 1986).
\bibitem{1'}
A. Peres and D.R. Terno, Rev. Mod. Phys. 76, 93 (2004).
\bibitem{1}
A. Einstein, B. Podolsky and N. Rosen, Phys. Rev. 47, 777 (1935).
\bibitem{2'}
M. Czachor, Phys. Rev. A 55, 72 (1997).
\bibitem{peres'}
A. Peres, P. F. Scudo, and D. R. Terno, Phys. Rev. Lett. 88, 230402,
(2002).
\bibitem{4}
P.M. Alsing and G. Milburne, Quant. Inf. Comp. 2, 487 (2002).
\bibitem{ref4}
R.M. Gingrich and C. Adami, Phys. Rev. Lett. 89, 270402 (2002).
\bibitem{5'}
D. Ahn, H.J. Lee, Y.H. Moon, and S.W. Hwang, Phys. Rev. A 67,
012103 (2003).
\bibitem{6'}
D. Ahn, H.J. Lee, and S.W. Hwang, e-print quant-ph:/0207018
(2002); D. Ahn, H.J. Lee, S.W. Hwang, and M.S. Kim, e-print
quant-ph:/0207018 (2003).
\bibitem{7'}
J. Pachos and E. Solano, Quant. Inf. Comp. 3, 115 (2003).
\bibitem{8'}
H. Terashima and M. Ueda, Int. J. Quant. Info. 1, 93 (2003).
\bibitem{9'}
A.J. Bergou, R.M. Gingrich, and C. Adami, Phys. Rev. A 68, 042102
(2003).
\bibitem{10'}
C. Soo and C.C.Y. Lin, Int. J. Quant. Info. 2, 183 (2003).
\bibitem{11'}
W.T. Kim, E.J. Son, e-print quant-ph:/0408127 (2004).
\bibitem{13'}
A. Peres, P. F. Scudo and D. R. Terno, Phys. Rev. Lett. 88 (2002)
230402.
\bibitem{14'}
A. Peres and D. R. Terno, Int. J. Quantum. Inform. 1, 225 (2003).
\bibitem{15'}
P. M. Alsing and G. Milburn, Phys. Rev. Lett. 91, 180404 (2003).
\bibitem{16'}
M. Czachor and M. Wilczewski, Phys. Rev. A 68, 010302 (2003).
\bibitem{17'}
H. Terashima and M. Ueda, Einstein-Podolsky-Rosen correlation in the
gravitational field, quant-ph/0307114.
\bibitem{18'}
R. Laiho, S. N. Molotkov and S. S. Nazin, Phys. Lett. A275 (2000)
36.
\bibitem{19'}
J. Rembielinski and K. A. Smolinski, Phys. Rev. A  66, 052114
(2002).
\bibitem{witbel}
P. Hyllus, O. G\"{u}hne, D. Bru{\ss}, and M. Lewenstein, Phys. Rev.
A 72, 012321 (2005).
\bibitem{witmeas}
Fernando G. S. L. Brand\"{a}o, Phys. Rev. A 72, 022310 (2005).
\bibitem{hilbersh}
R. A. Bertlmann, K. Durstberger, B. C. Hiesmayr, and Ph. Krammer,
Phys. Rev. A 72, 052331 (2005).
\bibitem{16} R. A. Bertlmann, H. Narnhofer, and W. Thirring, Phys. Rev. A
66, 032319 (2002).
\bibitem{Lewen3}
A. Ac\'{i}n, D. Bru\ss, M. Lewenstein,  and A. Sanpera, Phys. Rev.
Lett. \textbf{87}, 040401 (2001).
\bibitem{reza}
M. A. Jafarizadeh, M. Rezaee and S. Ahadpour, Phys. Rev. A 74,
042335 (2006).
\bibitem{reza1}
M. A. Jafarizadeh, M. Rezaee and S. K. A. Seyed Yagoobi, Phys.
Rev. A. 72, 062106 (2005).
\bibitem{Boyd}
S. Boyd and L. Vandenberghe, \emph{Convex Optimization}, Cambridge
University Press, (2004).
\bibitem{jsn}
M. A. Jafarizadeh, G. Najarbashi and H. Habibian, Phys. Rev. A,
75, 052326 (2007).
\bibitem{pachos}
J. Pachos and E. Solano, eprint: quant-ph/0203065.
\bibitem{woro}
S. L. Woronowicz, Rep. Math. Phys. 10, 165 (1976).
\bibitem{20}
M. Lewenstein,Quantum Information Theory, Winter Semester 2000/2001,
Institute for Theoreticl Physics, University of Hannover, March 31,
2004 (www.itp.uni-hannover.de/tqowww/ download/QIT2000.pdf).
\bibitem{29}
W. Dur and J. I. Cirac, Phys. Rev. A 61, 042314 (2000).
\bibitem{Horod2}
M. Horodecki, P. Horodecki, and R. Horodecki, Phys. Lett. A
\textbf{223}, 1 (1996).
\bibitem{Doherty3}
R. O. Vianna, A. C. Doherty, eprint quant-ph/0608095 (2006).
\bibitem{ohnuki}
Y. Ohnuki, \it{Unitary representations of the Poincare group and
relativistic wave equations}, Nagoya University, 1988.
\bibitem{Weinberg}
S. Weinberg, \it{The quantum theory of fields I}, Cambridge
University Press, N.Y. (1995).
\bibitem{woot}
W. K. Wootters, Phys. Rev. Lett. 80 2245 (1998).
\bibitem{horo}
M. Horodecki, P. Horodecki, and R. Horodecki, Phys. Rev. Lett. 80,
5239 (1998).
\end{thebibliography}
\end{document}